 %%%%%%%This requires the PHYZZX.TEX macropackage

\tolerance=10000
\input phyzzx

%%%%%%%If you do not have the msbm fonts, delete the following 4 lines
\font\mybb=msbm10 at 12pt
\def\bbbb#1{\hbox{\mybb#1}}
\def\Z {\bbbb{Z}}
\def\R {\bbbb{R}}
%%%%%%%%%%%%
%%%and replace with the following 2 lines (without %)
%\def\Z {Z}
%\def\R {R}
%%%%%%%%%%

%%%%%%%%%%%%%%%%%%%%%%%%%%%%%%%%%

%%%%%%%%%%%%
%%%%%%%%%%%%
%\makeatletter
%\new@fontshape{msb}{m}{n}{%
%   <5>msbm5%
%%   <6>msbm6%
 %  <7>msbm7%
 %  <8>msbm8%
 %  <9>msbm9%
 %  <10>msbm10%
 %  <11>msbm10%
 %  <12>msbm10 at10.95pt%
 %  <14>msbm10 at14.4pt%
 %  <17>msbm10 at17.28pt%
 %  <20>msbm10 at20.74pt%
 %  <25>msbm10 at24.88pt}{}
%\extra@def{msb}{}{}%{\noaccents@}
%\newmathalphabet*\bbl{msb}{m}{n}
%\makeatother

 %% semi-direct product
	\def \aa {\alpha}
\def \bb {\beta}
\def \gg {\gamma}
\def \dd {\delta}
\def \ee {\epsilon}

\def \ll {\lambda}
\def \mm {\mu}
\def \nn {\nu}

\def \ss {\sigma}
\def \tt {\tau}

 \def \ggg {\Gamma}

\def \eee {\varepsilon}

\def \www{\Omega}

\def\sym {super Yang-Mills}
\def \ti {\tilde}

\def \2 {{1 \over 2}}
\def \3 {{1 \over 3}}
\def \4 {{1 \over 4}}
\def \5 {{1 \over 5}}
\def \6 {{1 \over 6}}
\def \7 {{1 \over 7}}
\def \8 {{1 \over 8}}
\def \9 {{1 \over 9}}
\def \0 { \infty}

\def\++ {{(+)}}
\def \- {{(-)}}
\def\+-{{(\pm)}}

\def\ek {\eqn\abc$$}

\def \pa {\partial}

\def \qq {\qquad}

%%%%%%%%%%%%%%%%%%%%%%%%%%%%%%%%%%%%%%%%%%%%%%%%%%%%%%%%%%%%%%%%%%%%

 \def\unit{\hbox to 3.3pt{\hskip1.3pt \vrule height 7pt width .4pt \hskip.7pt
\vrule height 7.85pt width .4pt \kern-2.4pt
\hrulefill \kern-3pt
\raise 4pt\hbox{\char'40}}}

\def\nup#1({Nucl.\ Phys.\  {\bf B#1}\ (}

%%%%%%%%%%%%%%%%%%%%%%%%%
\REF\moore{G. Moore, hep-th/9305139,9308052.}
\REF\hsta{C.M. Hull, J.High Energy Phys. {\bf 7}, (1998) 021; hep-th/9806146.}
\REF\huku{C.M. Hull and R.R. Khuri, hep-th/9808069.}
\REF\mal{J. Maldacena,   hep-th/9711200.}
\REF\gat{E. Bergshoeff and E.  Sezgin,  Phys.  Lett. {\bf B292}  (1992) 87;
S.V. Ketov, H. Nishino and  S.J. Gates, Phys. Lett. {\bf B297}  (1992) 99,
Phys. Lett. {\bf B307} ( 1993)  323 
and Nucl.Phys. {\bf B393} (1993)  149.
}
\REF\duf{M.P. Blencowe and  M.J. Duff, 
Nucl.Phys.B310:387,1988. }
\REF\gibras{G.W. Gibbons and D.A. Rasheed, hep-th/904177.}
\REF\Cham{A. Chamblin and R. Empran, hep-th/9607236.}
\REF\bor{P. Goddard, \lq The Work of R.E. Borcherds', to appear in {\it Proceedings of the International Congress of Mathematicians}, and references therein.}
\REF\ddua {J. Dai, R.G. Leigh and J. Polchinski, Mod. Phys. Lett. {\bf
A4} (1989) 2073.}
\REF\dsei{ M. Dine, P. Huet and N. Seiberg, Nucl. Phys. {\bf B322}
(1989) 301.}
\REF\HJ{C. M. Hull  and B. Julia, hep-th/9803239.}
\REF\CPS{
E. Cremmer,  I.V. Lavrinenko,  H. Lu,  C.N. Pope,  K.S. Stelle and  T.A. Tran, hep-th/9803259.} 
\REF\FMS{D. Friedan, E. Martinec and S. Shenker, Nucl. Phys. {\bf B271} (1986) 93.}
\REF\polch{S. Chaudhuri, C. Johnson, and J. Polchinski,
``Notes on D-branes,'' hep-th/9602052; J. Polchinski,
``TASI Lectures on D-branes,'' hep-th/9611050; C.P. Bachas, hep-th/9806199.}
\REF\asp{P. Aspinwall, Nucl. Phys. 
Proc. Suppl. {\bf  46}  (1996) 30, hep-th/9508154; J. H. Schwarz, hep-th/9508143.}
\REF\ythe{C.M. Hull, Nucl.Phys. {\bf B468}  (1996) 113  hep-th/9512181.}
\REF\bergort{E. Bergshoeff, C.M. Hull and T. Ortin, Nucl. Phys. {\bf B451} (1995) 547, hep-th/9504081.}
\REF\gibrap{G.W. Gibbons, hep-th/9803206.}
\REF\buscher{ T. H. Buscher, Phys. Lett. {\bf 159B} (1985) 127, Phys. Lett. {\bf B194}
 (1987), 51 ; Phys. Lett. {\bf B201}
 (1988), 466.}
\REF\rocver {M. Ro\v cek and E. Verlinde, Nucl. Phys.
{\bf B373} (1992), 630.} 
 \REF\givroc {A. Giveon, M.
Ro\v cek, Nucl. Phys. {\bf B380} (1992), 128.}
\REF\alv{E. Alvarez, L. Alvarez-Gaum\' e, J.L. Barbon and Y. Lozano,  Nucl. Phys. {\bf B415} (1994)
71.}
\REF\TD {A. Giveon, M. Porrati and E. Rabinovici, Phys. Rep. {\bf 244}
(1994) 77.}
\REF\HS{C.M. Hull and B. Spence, Phys. Lett. {\bf 232B} (1989) 204.} 
\REF\wittz{E. Witten, Nucl. Phys. {\bf B443} (1995)  85, hep-th/9503124.}
\REF\CJ{E. Cremmer and B. Julia, Phys. Lett. {\bf 80B} (1978) 48; Nucl.
Phys. {\bf B159} (1979) 141.}
\REF\fhrs{W. Fischler, E. Halyo, A. Rajaraman and L. Susskind,  
hep-th/9703102.}
\REF\KT{T. Kugo and P.K. Townsend, Nucl. Phys. {\bf B221} (1983) 357.}
\REF\vann{P. van Nieuwenhuizen, Trieste lectures, 1981.}
\REF\HT{C.M. Hull and P.K. Townsend, hep-th/9410167.}
\REF\huten{C.M. Hull, Phys. Lett. {\bf B357 } (1995) 545, hep-th/9506194.}
\REF\dab{A. Dabholkar, Phys. Lett. {\bf B357} (1995) 307, hep-th/9506160.}
\REF\witpol{J. Polchinski and E. Witten,   hep-th/9510169.}
\REF\horwit{P. Horava and E. Witten, Nucl.Phys. {\bf B475} (1996) 94.}
\REF\fvaf{C. Vafa, Nucl. Phys. {\bf 469} (1996) 403.}

%%%%%%%%%%%%%%%%%%%%%%%%%%%%%%%%%%%%%%%%%%%%%%%%%%%%%%%%%%%%%%%%%%%%

%%%%%%%%%%%%%%%%%%%%%%%%%%%%%%%%%%%%%%%%%%%%%%%%%%%%%%%%%%%%%%%%%%%%
%%%%%%%%%%%%%%%%%%%%%%%%%%%%%%%%%%%%%%%%%%%%%%%%%%%%%%%%%%%%%%%%%%%%
%%%%%%%%%%%%%%%%%%%%%%%%%%%%%%%%%%%%%%%%%%%%%%%%%%%%%%%%%%%%%%%%%%%%

%%%%%%%%%%%%%%%%%%%%%%%%%%%%%%%%%%%%%%%%%%%%%%%%%%%%%%%%%%%%%%%%%%%%

%%%%%%%%%%%%%%%%%%%%%%%%%%%%%%%%%%%%%%%%%%%%%%%%%%%%%%%%%%%%%%%%%%%%

\Pubnum{ \vbox{  \hbox {QMW-PH-98-30} 
\hbox{hep-th/9807127}} }
\pubtype{}
\date{  July  1998}

\titlepage

\title {\bf     Duality and the Signature of Space-Time   
 }

\author{C.M. Hull}
\address{Physics Department, Queen Mary and Westfield College,
\break Mile End Road, London E1 4NS, U.K.}
\vskip 0.5cm

\abstract { Versions of M-theory are found in spacetime signatures (9,2) and (6,5), 
in addition to the usual M-theory in 10+1 dimensions, and these give rise to type IIA string 
theories in 10-dimensional spacetime signatures (10,0),(9,1),(8,2),(6,4) and (5,5), and to 
type IIB string 
theories in   signatures  (9,1),(7,3)  and (5,5).   The field theory limits are  10 and 11 dimensional supergravities in
these signatures. These theories are all linked by duality transformations which can change the number of time
dimensions as well as the number of space dimensions, so that each should be a different limit of the same underlying theory. }

\endpage

%%%%%%%%%%%%%%%%%%%%%%%%%%%%%%%%
%
% S-Tables Macro
%
%\message{S-Tables Macro v1.0, ACS, TAMU (RANHELP@VENUS.TAMU.EDU)}
%
% Help Text
%
\newhelp\stablestylehelp{You must choose a style between 0 and 3.}%
\newhelp\stablelinehelp{You should not use special hrules when stretching
a table.}%
\newhelp\stablesmultiplehelp{You have tried to place an S-Table inside another
S-Table.  I would recommend not going on.}%
%
% Line Thicknesses (Values)
%
\newdimen\stablesthinline
\stablesthinline=0.4pt
\newdimen\stablesthickline
\stablesthickline=1pt
%
% Border and Internal Line Thicknesses
%
\newif\ifstablesborderthin
\stablesborderthinfalse
\newif\ifstablesinternalthin
\stablesinternalthintrue
\newif\ifstablesomit
\newif\ifstablemode
\newif\ifstablesright
\stablesrightfalse
%
% Save Registers
%
\newdimen\stablesbaselineskip
\newdimen\stableslineskip
\newdimen\stableslineskiplimit
%
% Counters
%
\newcount\stablesmode
\newcount\stableslines
\newcount\stablestemp
\stablestemp=3
\newcount\stablescount
\stablescount=0
\newcount\stableslinet
\stableslinet=0
%
% Table Style Selection
%
% 0 - Centered
% 1 - Left Justified
% 2 - Right Justified
% 3 - Not Justified
%
\newcount\stablestyle
\stablestyle=0
%
% Element Buffering Definitions
%
\def\stablesleft{\quad\hfil}%
\def\stablesright{\hfil\quad}%
%
% Vertical Bar Activation
%
\catcode`\|=\active%
%
% Strut Control
%
\newcount\stablestrutsize
\newbox\stablestrutbox
\setbox\stablestrutbox=\hbox{\vrule height10pt depth5pt width0pt}
\def\stablestrut{\relax\ifmmode%
                         \copy\stablestrutbox%
                       \else%
                         \unhcopy\stablestrutbox%
                       \fi}%
%
% Misc. Internal Stuff
%
\newdimen\stablesborderwidth
\newdimen\stablesinternalwidth
\newdimen\stablesdummy
\newcount\stablesdummyc
\newif\ifstablesin
\stablesinfalse
%
% Table Macros
%
\def\begintable{\stablestart%
  \stablemodetrue%
  \stablesadj%
  \halign%
  \stablesdef}%
\def\stablesadj{%
  \ifcase\stablestyle%
    \hbox to \hsize\bgroup\hss\vbox\bgroup%
  \or%
    \hbox to \hsize\bgroup\vbox\bgroup%
  \or%
    \hbox to \hsize\bgroup\hss\vbox\bgroup%
  \or%
    \hbox\bgroup\vbox\bgroup%
  \else%
    \errhelp=\stablestylehelp%
    \errmessage{Invalid style selected, using default}%
    \hbox to \hsize\bgroup\hss\vbox\bgroup%
  \fi}%
\def\stablesend{\egroup%
  \ifcase\stablestyle%
    \hss\egroup%
  \or%
    \hss\egroup%
  \or%
    \egroup%
  \or%
    \egroup%
  \else%
    \hss\egroup%
  \fi}%
\def\stablestart{%
  \ifstablesin%
    \errhelp=\stablesmultiplehelp%
    \errmessage{An S-Table cannot be placed within an S-Table!}%
  \fi
  \global\stablesintrue%
  \global\advance\stablescount by 1%
  \message{<S-Tables Generating Table \number\stablescount}%
  \begingroup%
  \stablestrutsize=\ht\stablestrutbox%
  \advance\stablestrutsize by \dp\stablestrutbox%
  \ifstablesborderthin%
    \stablesborderwidth=\stablesthinline%
  \else%
    \stablesborderwidth=\stablesthickline%
  \fi%
  \ifstablesinternalthin%
    \stablesinternalwidth=\stablesthinline%
  \else%
    \stablesinternalwidth=\stablesthickline%
  \fi%
  \tabskip=0pt%
  \stablesbaselineskip=\baselineskip%
  \stableslineskip=\lineskip%
  \stableslineskiplimit=\lineskiplimit%
  \offinterlineskip%
  \def\borderrule{\vrule width \stablesborderwidth}%
  \def\internalrule{\vrule width \stablesinternalwidth}%
  \def\thinline{\noalign{\hrule height \stablesthinline}}%
  \def\thickline{\noalign{\hrule height \stablesthickline}}%
  \def\trule{\omit\leaders\hrule height \stablesthinline\hfill}%
  \def\ttrule{\omit\leaders\hrule height \stablesthickline\hfill}%
  \def\tttrule##1{\omit\leaders\hrule height ##1\hfill}%
  \def\stablesel{&\omit\global\stablesmode=0%
    \global\advance\stableslines by 1\borderrule\hfil\cr}%
  \def\el{\stablesel&}%
  \def\elt{\stablesel\thinline&}%
  \def\eltt{\stablesel\thickline&}%
  \def\elttt##1{\stablesel\noalign{\hrule height ##1}&}%
  \def\elspec{&\omit\hfil\borderrule\cr\omit\borderrule&%
              \ifstablemode%
              \else%
                \errhelp=\stablelinehelp%
                \errmessage{Special ruling will not display properly}%
              \fi}%
  \def\stmultispan##1{\mscount=##1 \loop\ifnum\mscount>3 \stspan\repeat}%
  \def\stspan{\span\omit \advance\mscount by -1}%
  \def\multicolumn##1{\omit\multiply\stablestemp by ##1%
     \stmultispan{\stablestemp}%
     \advance\stablesmode by ##1%
     \advance\stablesmode by -1%
     \stablestemp=3}%
  \def\multirow##1{\stablesdummyc=##1\parindent=0pt\setbox0\hbox\bgroup%
    \aftergroup\emultirow\let\temp=}
  \def\emultirow{\setbox1\vbox to\stablesdummyc\stablestrutsize%
    {\hsize\wd0\vfil\box0\vfil}%
    \ht1=\ht\stablestrutbox%
    \dp1=\dp\stablestrutbox%
    \box1}%
  \def\stpar##1{\vtop\bgroup\hsize ##1%
     \baselineskip=\stablesbaselineskip%
     \lineskip=\stableslineskip%
     \lineskiplimit=\stableslineskiplimit\bgroup\aftergroup\estpar\let\temp=}%
  \def\estpar{\vskip 6pt\egroup}%
  \def\stparrow##1##2{\stablesdummy=##2%
     \setbox0=\vtop to ##1\stablestrutsize\bgroup%
     \hsize\stablesdummy%
     \baselineskip=\stablesbaselineskip%
     \lineskip=\stableslineskip%
     \lineskiplimit=\stableslineskiplimit%
     \bgroup\vfil\aftergroup\estparrow%
     \let\temp=}%
  \def\estparrow{\vfil\egroup%
     \ht0=\ht\stablestrutbox%
     \dp0=\dp\stablestrutbox%
     \wd0=\stablesdummy%
     \box0}%
  \def|{\global\advance\stablesmode by 1&&&}%
  \def\|{\global\advance\stablesmode by 1&\omit\vrule width 0pt%
         \hfil&&}%
  \def\vt{\global\advance\stablesmode by 1&\omit\vrule width \stablesthinline%
          \hfil&&}%
  \def\vtt{\global\advance\stablesmode by 1&\omit\vrule width
\stablesthickline%
          \hfil&&}%
  \def\vttt##1{\global\advance\stablesmode by 1&\omit\vrule width ##1%
          \hfil&&}%
  \def\vtr{\global\advance\stablesmode by 1&\omit\hfil\vrule width%
           \stablesthinline&&}%
  \def\vttr{\global\advance\stablesmode by 1&\omit\hfil\vrule width%
            \stablesthickline&&}%
  \def\vtttr##1{\global\advance\stablesmode by 1&\omit\hfil\vrule width ##1&&}%
  \stableslines=0%
  \stablesomitfalse}
\def\stablesdef{\bgroup\stablestrut\borderrule##\tabskip=0pt plus 1fil%
  &\stablesleft##\stablesright%
  &##\ifstablesright\hfill\fi\internalrule\ifstablesright\else\hfill\fi%
  \tabskip 0pt&&##\hfil\tabskip=0pt plus 1fil%
  &\stablesleft##\stablesright%
  &##\ifstablesright\hfill\fi\internalrule\ifstablesright\else\hfill\fi%
  \tabskip=0pt\cr%
  \ifstablesborderthin%
    \thinline%
  \else%
    \thickline%
  \fi&%
}%
\def\endtable{\advance\stableslines by 1\advance\stablesmode by 1%
   \message{- Rows: \number\stableslines, Columns:  \number\stablesmode>}%
   \stablesel%
   \ifstablesborderthin%
     \thinline%
   \else%
     \thickline%
   \fi%
   \egroup\stablesend%
\endgroup%
\global\stablesinfalse}
%
% end of STABLES.TEX
%
\chapter{Introduction}

In recent years, it has been discovered that dualities  can relate theories with different gauge
groups, different space-time dimensions, different amounts of supersymmetry, and even relate
theories of gravity to gauge theories.
Thus many of the concepts that had been thought absolute are now understood as relative: they
depend on the \lq frame of reference' used, where the concept of frame of reference is generalised to
include the values of the various coupling constants. For example, the description of a given
system when a certain coupling is weak can be very different from the description at strong
coupling, and the two regimes can have different spacetime dimension, for example.
However, in all this, one thing that has remained unchanged is the number of time dimensions; all
the theories considered are formulated in a Lorentzian signature with one time coordinate, although
the number of spatial dimensions can change.
In this paper, it will be argued that dualities can change the number of time dimensions as well,
giving rise to exotic spacetime signatures.

The strong coupling limit of the type IIA superstring is a theory in 10+1 dimensions whose low
energy limit is 11-dimensional supergravity theory  and which is referred to as M-theory.
The type I, type II and heterotic superstring theories and 
certain supersymmetric gauge theories   emerge as different limits of
M-theory.
The M-theory in 10+1 dimensions will be linked via dualities to a theory in 9+2 dimensions (9 space and 2 time dimensions),
which will be referred to as $M^*$ theory, and a theory in 6+5 dimensions, $M'$-theory.
Various limits of these will give rise to IIA-like string theories in 10+0, 9+1,8+2,6+4 and 5+5
dimensions, and to IIB-like string theories in   9+1,7+3,  and 5+5
dimensions.
The field theory limits  are supergravity theories with 32 supersymmetries in  10 and 11 dimensions with these signatures,
many of which are new.
Further dualities relate these 
  to supersymmetric gauge theories in various signatures and dimensions, such as 2+2,3+1 and 4+0.

The resulting picture is that there should be some underlying fundamental theory and that different 
spacetime signatures as well as different dimensions can arise in various limits. 
The   new theories are different real forms of the complexification of the original M-theory
and type II string theories, perhaps suggesting an underlying complex nature of spacetime.

Each of the theories has a flat-space solution $\R^{p,q}$ of the appropriate signature, but they
are linked via   compactifications on tori $T^{m,n}$ with $m$ compact space dimensions and $n$ compact time dimensions.
The starting point is to note that  conventional string theory  and M-theory with Lorentzian
signature have  classical solutions which are flat spaces of
 the form $\R^m\times T^{n,1}$ where 
$T^{n,1}$ is a Lorentzian torus with $n$ spacelike circles and one timelike circle.
The presence of closed timelike loops means that the physics in such spaces is unusual, but
 it has often been fruitful in the past to 
study solutions that have little in common with the real world. An important issue with these
solutions (as with many others) is whether a consistent quantum theory can be formulated in such
backgrounds. 
Wave equations or Schr\" odinger equations can be solved with periodic time, but issues of measurement
and collapse of the wave-function are problematic. In string theory, it is straightforward to study
the solutions of the physical state conditions, but there are new issues that arise from strings
(and branes) winding around the compact time.

Assuming that such timelike   compactifications are consistent, then it is of interest to
investigate the \lq strong coupling' limits that arise in going to the boundaries of the moduli space
and, surprisingly, the resulting limits give unexpected new theories.
 A superstring theory in 9+1 dimensions compactified on a timelike circle gives, in the limit
in which the circle shrinks to zero size, another superstring theory in 9+1 dimensions related by
timelike T-duality. 
Timelike T-duality takes the heterotic string back to the heterotic string [\moore], but
it was shown in [\hsta] that  the timelike T-duals of the type IIA or type IIB
strings are
\lq new' theories, the type $IIB^*$
or type $IIA^*$ strings, respectively.
The strong coupling limit of the $IIA^*$ theory is the $M^*$-theory in 9+2 dimensions, as will be
shown in section 6. Whereas M-theory compactified on a Euclidean 3-torus $T^{3,0}$ gives  M-theory
again in the limit in which the torus shrinks to zero size, M-theory  
 compactified on a Lorentzian 3-torus $T^{2,1}$, in the
limit in which the torus shrinks to zero size,  gives the $M^*$ theory in 9+2 dimensions. 
Next
compactifying $M^*$ theory on a Euclidean 3-torus $T^{3,0}$ gives, in the
  zero size limit, the $M'$ theory in 6+5 dimensions.
Compactifying the $M,M^*,M'$ theories on   circles or 2-tori of various signatures and taking the zero volume limit
then give rise to
the various 10-dimensional string theories.
These 11 and 10-dimensional theories have   brane solutions with various world-volume signatures
[\hsta,\huku] and    some of these
interpolate between flat space and  solutions which are the product of a symmetric space
$SO(p,q)/SO(p,q-1)$, which is a generalised de Sitter or anti de Sitter space of signature $(p,q-1)$,
with an internal hyperbolic or spherical symmetric space.
The world-volume theory is a superconformal  gauge theory in a spacetime with signature
$(p-1,q-1)$ which is invariant under the conformal group $SO(p,q)$. This leads to a duality between
the superconformal gauge theory and the string or M-theory in the   de Sitter-type space [\hsta], by
arguments similar to those in [\mal].

It has been known for some time that there are supersymmetric theories in non-Lorentzian
signatures, such as 2+2 or 5+5 [\gat],  and branes of various signatures were
discussed in [\duf].
 The considerations here lead to new supergravity theories to
add to this list, and 
the string or M-theories  that lead to these are all related to one another and to the
supersymmetric gauge theories in various signatures by dualities, thereby unifying the many
maximally supersymmetric
 theories in diverse signatures. The web of dualities linking these theories has a robust structure
which survives many consistency checks, and it is remarkable that 
the theories that arise are all consistent with supersymmetry; for example, IIB theories arise in
precisely those signatures admitting chiral fermions and real self-dual 5-forms.

Many of the theories that arise in this way  appear to have
 pathologies, such as ghosts, tachyons or
instabilities. 
Most of the theories have instantons [\hsta,\huku,\gibras,\Cham] that may lead to the instability of flat space, but,
 at least in some cases, there is an alternative interpretation of these solutions [\hsta] and the question of vacuum decay is
unclear.

To summarise, M-theory and superstring theory have classical solutions that have periodic time.
If a consistent quantum string or M-theory exists in such backgrounds, then
dualities lead to the new theories with exotic signatures that are the subject of this paper, and 
 the fact that the new theories are the usual M-theory or
string theory written in terms of   dual variables would mean that 
they are no worse than the original theory (with compact time) and
some of the features that appear to be
pathological are in fact
the consequence of using unusual variables to describe the theory; this will be discussed further
in section 3. It would of course be important and interesting to understand how   physics might work  in such theories directly, without recourse to duality arguments. These theories in signature $(s,t)$ also have decompactification  limits which are flat solutions $\R^{s,t}$,
and the question arises as to whether the theories in these backgrounds are consistent or have some pathology. For example, 
although the various  \lq wrong' signs occuring in the theories suggest instabilities, in some cases at least 
  some of the apparent instabilities appear to be absent [\hsta].
Of course, it may be that 
M-theory or string theory with periodic time is in fact pathological and then the new 
  theories related to these by duality would also be
  pathological, and  these pathologies would
have been introduced  by introducing closed timelike curves.
However, the new theories can be written down in flat space without any reference to timelike
curves, and the timelike compactifications are used only to relate the theories one to   another.
(It is perhaps worth mentioning that string theory in backgrounds in which time as well as all the space dimensions are compact have played an important role in the study of vertex algebras, the monster group and related areas of mathematics; see e.g. [\moore,\bor].)
It is an important feature that all the exotic new theories are linked to the usual M-theory by chains of dualities, so that there should  still be a single underlying theory.
However, it seems that M-theory, or whatever the underlying theory is to be called, may have \lq phases'
in which the physics is    more unusual than previously suspected.

\chapter{The Type $II^*$ Superstring Theories}

T-duality on a space-like circle interchanges the  IIA and IIB  string theories  [\ddua,\dsei], but 
for a timelike circle, T-duality  does not take the  IIA string theory to a IIB  string theory or
the IIB to IIA [\CPS,\hsta]. In [\hsta], the images of the IIA and IIB theories under timelike T-duality,
the 
 $IIB^*$ and $IIA^*$   theories respectively,  were investigated.  The $IIA^*$ and $IIB^*$ theories
can be obtained from the IIA and IIB  theories, respectively,  by acting with $i^{F_L}$ where $F_L$ is the left-handed
fermion number. The zero-slope limits give corresponding supergravity actions, which differ from the
usual IIA and IIB supergravities by certain signs; in particular, the signs of the kinetic terms of
the RR gauge fields are all reversed, and this leads to the presence of E-brane solutions instead of
the D-branes of the type II theories. An E$n$-brane  arises from imposing   Dirichlet boundary
conditions in time as well as in
$9-n$ spatial 
  coordinates, and 
is associated with 
an extended object with $n$ space-like dimensions occurring at a particular instant in time.
The E$n$-branes of the type $II^*$ theories are related to the D$n$-branes of the type II theories by a
timelike T-duality. The T-dualities linking the type II and type $II^*$ theories are illustrated in the following diagram.

\vskip .5 cm

{\vbox{

%%Begin InstantTeX Picture
\let\picnaturalsize=N
\def\picsize{3.0in}
\def\picfilename{SigFigs/fig12.eps}
%If you do not have the picture file add:
%\let\nopictures=Y
%to the beginning of the file.
\ifx\nopictures Y\else{\ifx\epsfloaded Y\else\input epsf \fi
\let\epsfloaded=Y
\centerline{\ifx\picnaturalsize N\epsfxsize \picsize\fi \epsfbox{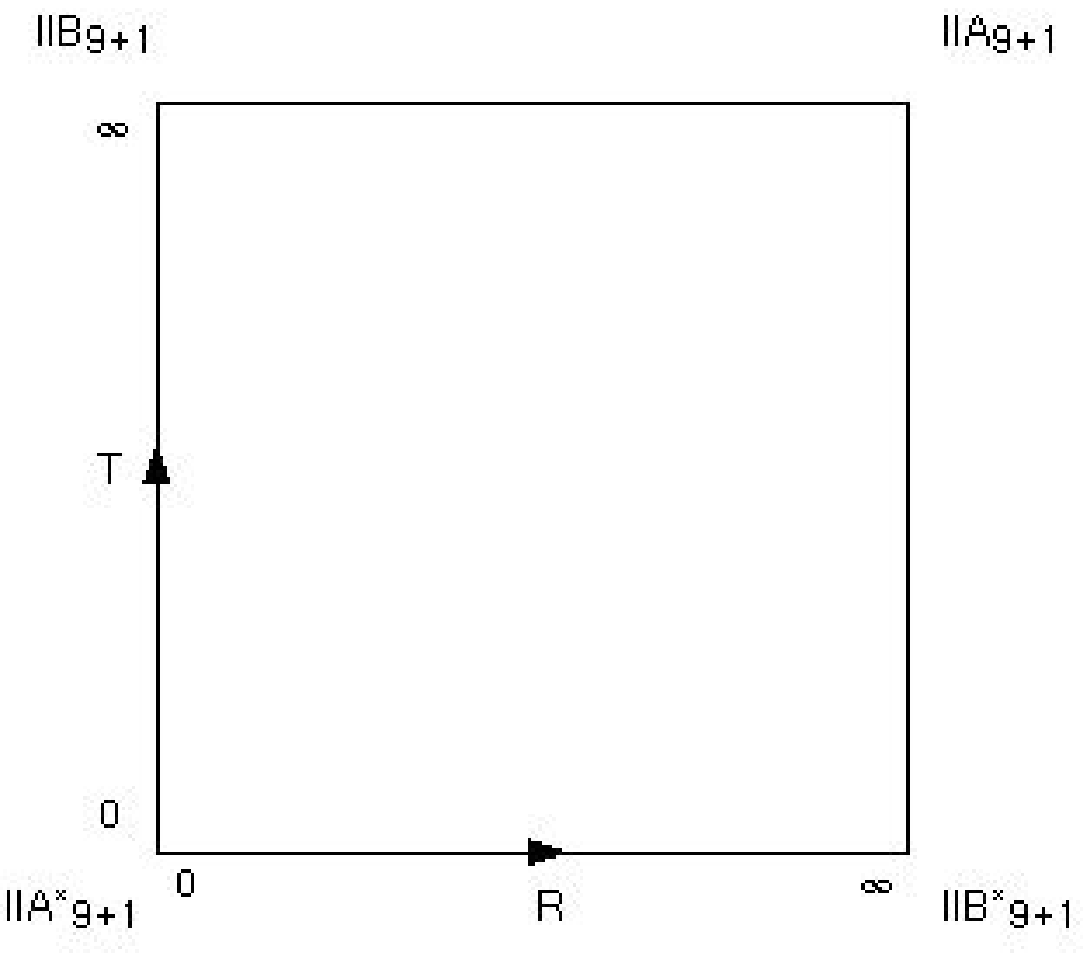}}}\fi
%%End InstantTeX Picture

{\bf Figure 1}  The moduli space for the type IIA string theory compactified
   on a  Lorentzian torus $T^{1,1}$  with  spacelike radius $R$ and  timelike radius  $T$.}}

\vskip .5 cm

\noindent A spacelike T-duality relates the IIA with the IIB theory and the $IIA^*$ with the $IIB^*$ theory, while a timelike
T-duality links the $IIA$ with the $IIB^*$ theory and the $IIB$ with the $IIA^*$ theory.

The type $II^*$ theories have a twisted $N=2$ 
superalgebra   
$$ \{Q_i,Q_j \} = \eta _{ij} (\ggg ^\mm C^{-1})P_\mm 
\eqn\salg$$
where $i,j=1,2$ labels the two supercharges (which have the same chirality in the $IIB^*$ theory
and opposite chirality in the $IIA^*$ theory), $C$ is the charge conjugation matrix 
 and  $ \eta _{ij}$
is the
$SO(1,1)$ invariant metric
$diag(1,-1)$.  
The anti-commutator of the second supercharge with itself has the \lq wrong' sign.
The scalars  of the $IIB^*$ theory take values in the coset space
$$
{SL(2,\R)\over SO(1,1)}\eqn\abc$$
 instead of the coset space
$$
{SL(2,\R)\over SO(2)}\eqn\abc$$
of the IIB theory.
The compactifications of these theories on spacelike tori $T^n$ and Lorentzian tori $T^{n,1}$ (with $n$ spacelike circles and
one timelike one) were considered in [\hsta], generalising the results of [\HJ,\CPS].
In particular, whereas the type IIB theory is obtained from M-theory by reduction on $T^2$ in the limit in which the torus
shrinks to zero size, the $IIB^*$ theory arises
from M-theory compactified on $T^{1,1}$ in the limit in which it shrinks to zero size.

Given the existence of these \lq new' type $II^*$ theories, it is natural to ask whether any further new theories can be
obtained in a similar way. This is indeed the case, and the purpose of this paper is to explore these new theories and the
web of dualities linking them to each other and to M-theory. We start by pointing out two apparent problems arising from the
findings of [\hsta], the resolution of which leads to the first of the new theories. 

First, M-theory compactified on a timelike circle gives a  string theory in 10 Euclidean dimensions 
  which we will refer to
as the
$IIA_E$ theory (or $IIA_{10+0}$ theory). If this string theory is compactified on a   circle, and the limit of zero radius is
taken, the result should be the T-dual of the $IIA_E$ theory, which should again be a 10-dimensional string theory. The naive
expectation is that it should again be in 10 Euclidean dimensions, and should be some Euclidean version of the type IIB
theory. However, this cannot be the case, for a number of reasons. First, the IIB theory has chiral fermions and a 4-form
gauge field with self-dual field strength, and there is no Euclidean version  that does not double some of these degrees of
freedom. Second, the supergravity limit of the T-dual of the $IIA_E$ theory should, when reduced to nine dimensions, give the
same  9-dimensional theory as dimensionally reducing the $IIA_E$ theory; 
this 9-dimensional Euclidean theory is the one referred to
as the
$IIA_9$ theory in [\hsta]. The only IIB-type supergravity theory in 10 dimensions compactifying to 
this theory is easily seen to be (up to field redefinitions) the 
$IIB^*$ theory compactified on a {\it timelike} circle, and the only Euclidean supergravity theory
with this reduction  (up to field redefinitions) is the  $IIA_E$ theory itself. 

Recall that the IIB string theory can be obtained by compactifying M-theory on a spacelike torus $T^2$ and taking the
limit in which the torus shrinks to zero size [\asp].
M-theory on a torus with radii $R_1,R_2$ can be represented in the following diagram:
\vskip .5 cm

{\vbox{

%%Begin InstantTeX Picture
\let\picnaturalsize=N
\def\picsize{3.0in}
\def\picfilename{SigFigs/fig1.eps}
%If you do not have the picture file add:
%\let\nopictures=Y
%to the beginning of the file.
\ifx\nopictures Y\else{\ifx\epsfloaded Y\else\input epsf \fi
\let\epsfloaded=Y
\centerline{\ifx\picnaturalsize N\epsfxsize \picsize\fi \epsfbox{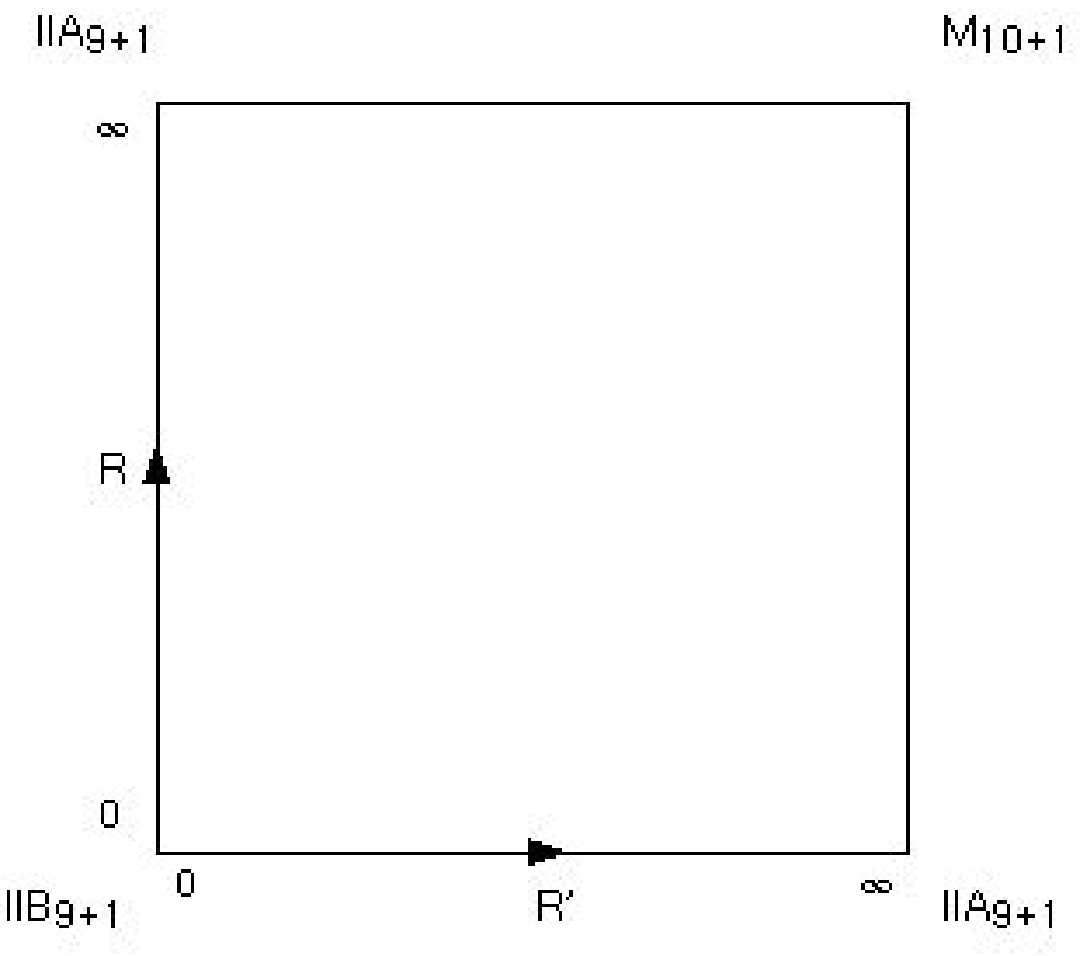}}}\fi
%%End InstantTeX Picture

%\vskip -.3 cm
 
{\bf Figure 2}  The moduli space for M-theory compactified on a rectangular torus $T^2$ with radii 
$R,R'$.  Points should be identified under the   symmetry  that interchanges   $R$ with $R'$.}}

\vskip .5 cm

\noindent If one of the radii shrinks, we get the weakly coupled IIA string and if both shrink, we obtain the IIB string with
string coupling given by the ratio $R_1/R_2$.

 Consider instead  M-theory compactified on the Lorentzian torus $T^{1,1}$. First
compactifying on a space-like circle gives type IIA string theory, and then  compactifying on a timelike circle and taking the
limit in which the circle shrinks to zero size gives the $IIB^*$ string theory, by timelike T-duality.
If now the order of reductions is reversed, the reduction on the timelike circle gives the $IIA_E$ theory and a further
reduction on a spacelike circle should again give the $IIB^*$ theory. This would lead one to   expect a situation represented
by the following diagram:

\vskip .5 cm

{\vbox{

%%Begin InstantTeX Picture
\let\picnaturalsize=N
\def\picsize{3.0in}
\def\picfilename{SigFigs/fig2.eps}
%If you do not have the picture file add:
%\let\nopictures=Y
%to the beginning of the file.
\ifx\nopictures Y\else{\ifx\epsfloaded Y\else\input epsf \fi
\let\epsfloaded=Y
\centerline{\ifx\picnaturalsize N\epsfxsize \picsize\fi \epsfbox{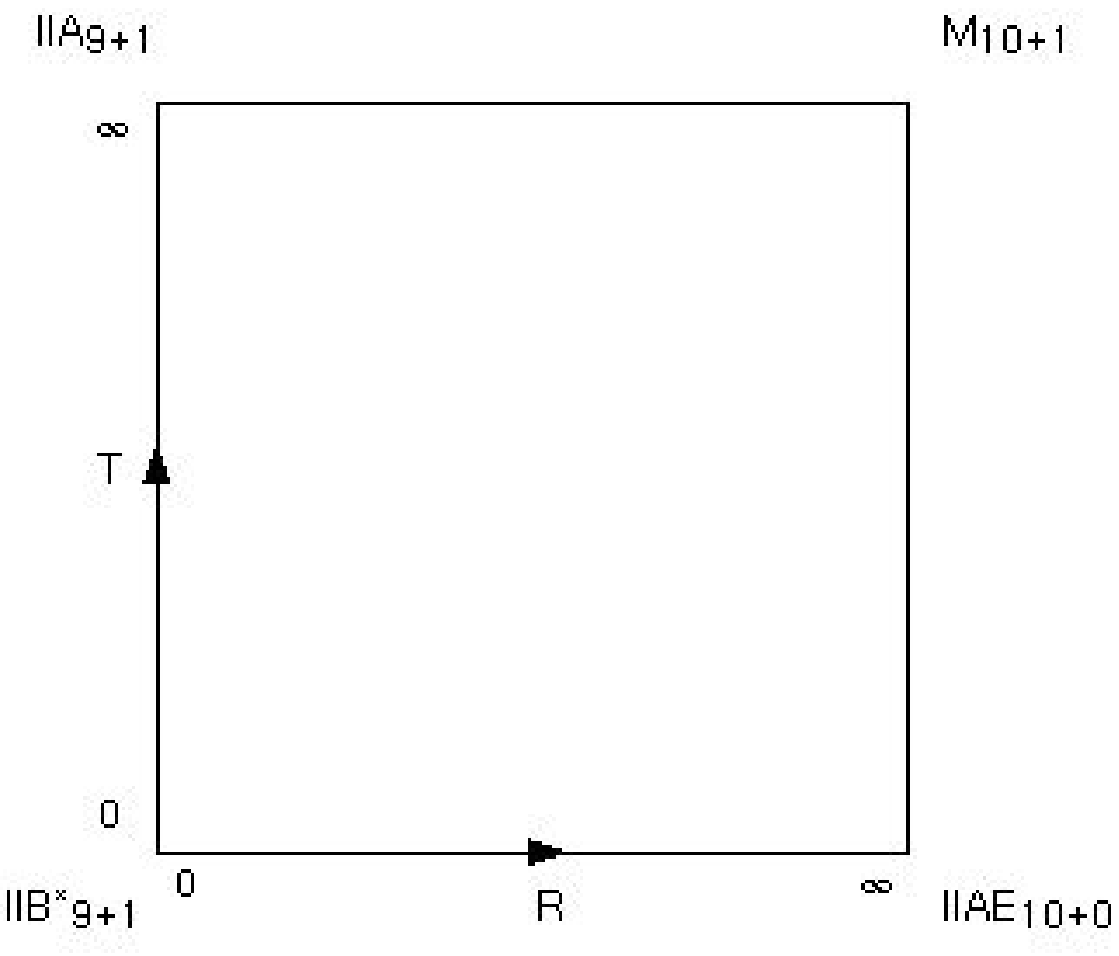}}}\fi
%%End InstantTeX Picture

%\vskip  -.3 cm

{\bf Figure 3}  The moduli space for M-theory compactified on a Lorentzian torus $T^{1,1}$ with spacelike radius $R$ and
timelike radius $T$. }}

\vskip .5 cm

\noindent Reducing on a $T^{1,1}$ with spacelike radius $R$ and timelike radius $T$ gives a moduli space  of theories labelled
by $R,T$ depicted in the diagram. If the order in which limits are taken does not matter, then the T-dual of the $IIA_E$
theory must be the $IIB^*$ theory. There are two obvious problems with this: (i) The T-duality would  change the
space-time signature, as it relates the $IIA_E$ theory   in 10 Euclidean dimensions to the $IIB^*$
theory    in 9+1 dimensions with Lorentzian signature. The $IIA_E$ theory on a {\it spacelike
circle} of radius $R$ is required to be T-dual to the $IIB^*$ theory on a {\it timelike circle} of
radius $1/R$.  (ii) The $IIB^*$ theory already has a T-dual, as the  $IIB^*$ theory on a { 
timelike circle} of radius $1/R$ is equivalent to the IIA theory on a { 
timelike circle} of radius $R$,
and the $IIB^*$ theory on a spacelike circle is T-dual to the  $IIA^*$ theory on a spacelike circle,
 so how can the  $IIB^*$ theory be T-dual to the $IIA_E$ theory as
well? 
On the other hand, if the order of the limits does matter, then the structure of the moduli
space near the point $R=T=0$ could be rather  complicated, as the different theories could arise,
depending on the direction in which this point was approached. Also, if the $IIB^*$ theory is not
the T-dual of the $IIA_E$ theory, then it is hard to find another candidate.
 These issues will be resolved in section 4 and 5, where   it will be argued that under certain
circumstances, T-duality {\it can} change a spacelike circle into a timelike one, and that the
$IIB^*$ theory on a timelike circle is indeed T-dual to both the $IIA_E$ and $IIA^*$ theories. 
The point is that the string coupling constant of the $IIB^*$ theory is proportional to $R/T$, so
that taking first $R$ small to obtain the weakly coupled IIA theory and then
taking $T$ small to give the timelike T-dual of IIA gives the $IIB^*$ at weak coupling, while
first taking $T$ small to give the weakly coupled  $IIA_E$ theory  and then taking $R$ small
to give its T-dual results in the strong coupling limit of the $IIB^*$ string theory.
The
strong coupling limit of the $IIB^*$ string theory is, as will be seen,  a {\it different} perturbative
string theory, the
$IIB'$ theory, although   the 
$IIB^*$ string theory
and the $IIB'$ string theory are different perturbative limits of the same underlying non-perturbative  theory.
The timelike T-dual of the $IIB^*$ string theory is the $IIA$ string theory, as expected, but the
timelike T-dual of the $IIB'$ string theory is the $IIA_E $ string theory.  However, 
having resolved this issue, we find the
T-dual of the
$IIB'$ theory on a space-like circle leads to something new again, as will be discussed in section 6.

A second open question is what is the strong coupling limit of the $IIA^*$ string theory. As in the IIA case, it
is expected to be some 11-dimensional theory. 
 However, it cannot have the usual 11-dimensional supergravity  as its field theory limit,
as the reduction of this gives the $IIA$ theory, not
the $IIA^*$  theory.
The natural guess is that it should have a 10+1 dimensional field theory
limit, but we will see that there is no such theory which gives the right reduction.
 This will be
addressed in section 7, and the strong coupling limit will be identified as a theory in $9+2$ dimensions, and  
a field theory in 9+2 dimensions which dimensionally reduces to the $IIA^*$ theory will be found.

\chapter{Timelike T-duality}

Consider first the bosonic or heterotic string.
For a flat string theory background which includes a  periodic coordinate $X$, the
most general string configuration in that direction $X(\ss,\tt)$ is   
$$X (\ss ,\tt) = x + p \tt + \ti p \ss + \sum_n \left( a_n  e^{in(\tt+\ss) }+
\ti a_n  e^{in(\tt-\ss) }\right)
\eqn\abc$$
where $\tt,\ss$ are the closed-string world-sheet coordinates,   with periodic   coordinate
$\ss \sim \ss + 2 \pi$, and $\tt \in \R$.
If $X$ is periodically identified 
$X \sim X + 2\pi R$ for some radius  $R$, then the
phase $ e^{ip\cdot x} $ will be single-valued if  the momentum $p $ is quantized
$$p ={ n\over R}
\eqn\abc$$
with $n $ an integer, while the dual momentum $\ti p$
must satisfy
$$\ti p ={ m  R}
\eqn\abc$$
 so that
$X$ is well-defined. Then $m$ is the number of times the  periodic coordinate $\ss$ of
the string world-sheet winds around the $X$ dimension.
The physical state conditions are invariant under the  T-duality transformation that interchanges $m$ and $n$
and takes $R\to 1/R$ and so
interchanges $p$ with $\ti p$. Writing 
$$X(\tt,\ss)= X(\tt+\ss)+\ti X(\tt -\ss )
\ek
the T-duality takes $X\to X$, and $\ti X \to - \ti X$.

In the above, the compact coordinate $X$ can be either spacelike or timelike, and if it is timelike,
$X=X^0$,  then $p=p^0$ is the energy, which is  quantized, $p^0=n/R$, and 
  $m $ is the number of times the $\ss$ coordinate on the world-sheet winds around the timelike direction in the
target space.
For a configuration in which the compact world-sheet   coordinate $\ss$ maps to the target space time $X^0$, the non-compact 
world-sheet   coordinate $\tt$ would map to a spatial direction $X^1$, say, in the target space,  if the world-sheet
is to be
 non-degenerate. 
Note that for such   configurations, $\tt$ becomes   a space coordinate and $\ss$ a periodic time coordinate with respect to
the induced world-sheet metric.
Then $p^\mm= \partial _\tt X^\mm$ is a tachyonic (spacelike) momentum, and it is straightforward to see
that such states must occur.
Consider, for example, a conventional string state with timelike momentum $(p^0=n/R,p^i)$ and no winding number.
Then a timelike T-duality would take this to a state with timelike winding number $n$ and momentum $(0,p^i)$, which is
spacelike and tachyonic.
Nonetheless, this is the original \lq good'  state   described in terms of dual variables, so that the unusual features
are a consequence of the variables used and do not reflect any pathology of the original state.
For the heterotic string compactified on time to $\R^9\times S^1$, both the limits $R\to \infty$ and $R \to 0$ give the
heterotic string in 9+1 dimensional Minkowski space which is free from both ghosts and  tachyons (at least perturbatively)
and is stable. For the tachyonic state described above, the winding number becomes the energy corresponding to the dual time
coordinate, and the resulting dual momentum is timelike, as it should be.
The perturbative spectrum for finite $R$ has been analysed in [\moore], where it was shown to be
free from ghosts, with a positive definite Hilbert space.

Consider the low-energy field theory, which includes
$$ 
S = 
\int
 d^{10} x  
\sqrt{-g} 
 e^{-2 \Phi} \left( R+ 4(\partial   \Phi )^2  
-H^2 
\right)+\dots
\eqn\het$$
where $\Phi$ is the dilaton and $H=dB_2$ is the field strength for the 2-form gauge field $B_2$.
Reducing this on time to 9 Euclidean dimensions gives an action 
$$ S = 
\int
 d^{9} x  
\sqrt{g}\left[
 e^{-2 \Phi} \left( R+ 4(\partial   \Phi )^2  
-H^2  - d \chi ^2 + F_g^2 +F_B^2
\right)+\dots \right]
\eqn\hetb$$
where $F_g,F_B$ are the field strengths for the one-form gauge fields 
$A_i^g \sim g_{i0},A_i^B \sim B_{i0}$ and $\chi$ is the scalar coming from the metric.
The T-duality leads to a symmetry of the 9-dimensional action under which $A^g,A^B$ are interchanged.
The kinetic terms for the two vector fields $A^g,A^B$ have the \lq wrong' sign, usually associated with ghosts.
However, if all the Kaluza-Klein modes are kept, the theory is the full 10-dimensional theory and the 10-dimensional gauge
symmetries can be used to remove all negative norm states, including   those corresponding to the fields
$g_{i0}, B_{i0}$, at least in the topologicaly trivial sector.
The 10-dimensional physical states   carry non-zero $p^0$ and all of these are thrown away in the truncation of dependence on
the $X^0$ coordinate, so that the dimensionally reduced action \hetb\ is particularly misleading for studying the physical
states.  Nonetheless, such actions can be useful for studying some aspects of the theory, and relations between theories, but
should be regarded, for most purposes,  as being accompanied by an action for an infinite set of   modes with non-zero energy,
and most of the physics is in these modes.

Consider now T-duality in the type II theories, which in the covariant NSR formalism are formulated in terms of 
bosonic coordinates $X^\mm$ and their   superpartners $\Psi ^\mm$, which are world-sheet spinors.
The transformation $\pa _a X ^\mm \to \ee _a {}^b \pa _b X^\mm$ or     $X^\mm+\ti X^\mm \to X^\mm - \ti X^\mm$ 
for T-duality in the $\mm$ direction is accompanied by the transformation   $\Psi ^\mm \to \gg ^3 \Psi ^\mm$.
Decomposing $\Psi ^\mm$ into a right-handed Majorana-Weyl world-sheet fermion $\psi^\mm$ and a left-handed one  $\ti \psi
^\mm$,  this transformation acts as 
$$\psi ^\mm \to \psi ^\mm, \qq \ti \psi ^\mm \to -\ti \psi ^\mm
\ek
In addition, there are right-handed and left-handed spin operators $S,\ti S$ which are Majorana space-time spinors satisfying
the chirality constraints
$\ggg^{11} S=S, \ggg^{11} \ti S= \ti S$ for the type IIB string and
$\ggg^{11} S=S, \ggg^{11} \ti S= -\ti S$ for the type IIA string [\FMS].
As a reflection in a spatial direction, $X^9$ say,  acts in spin space through the operator $i\ggg^{11}\ggg^9$,   T-duality
in the $X^9$ direction acts on the spin operators as [\polch]
$$S \to S, \qq \ti S \to  \ti S'=i\ggg^{11}\ggg^9\ti S
\eqn\spint$$
This changes the chirality of $\ti S$.
By considering the action on RR vertex operators [\polch],
this gives the relation between the RR field strengths $G_n'$ of the type IIA (IIB) theory with $n$ even (odd), and the
field strengths $G_{n\pm 1}$ of the T-dual type IIB (IIA) theory:
$$G'_{\mm_1 \dots \mm_n}=
G_{9\mm_1 \dots \mm_n}, \qq 
G'_{9\mm_1 \dots \mm_n}=-G_{\mm_1 \dots \mm_n}
\eqn\tytty$$
for any $\mm_i\ne 9$.

For a
timelike T-duality in the $X^0$ direction, 
the T-duality should take $\ti S \to P\ti S$ for some operator $P$ which should be given by $ \ggg^{11}\ggg^0$, up to a phase.
There are possible phases (up to a sign), which lead to equivalent results. The operator $P=i\ggg^{11}\ggg^0$ preserves the
Majorana condition on
$\ti S$ but is not unitary, while
$ \ggg^{11}\ggg^0$ is unitary but does not preserve the Majorana condition.
Choosing the unitary operator, 
then (in a Majorana representation in which Majorana spinors are real and the gamma matrices imaginary) $\ti S'$ is imaginary.
As a result,  a timelike T-duality 
maps the type II theory with   real (Majorana) spin operators $S,\ti S$ to a 
new theory, the type $II^*$ theory, in which there is a real right-handed spin operator
$S$   and an  imaginary left-handed  one $\hat S$,  with the T-duality taking 
  $\ti S \to \hat S = \ggg^{11}\ggg^0 S$.
If we choose instead  $P=i\ggg^{11}\ggg^0$, 
the T-duality preserves the reality of $\ti S$ and changes its chirality, 
but the fields in the R-R and NS-R sectors (i.e. the sectors constructed using $\ti S$) become imaginary, so that 
the timelike T-dual of the IIA (IIB) theory is given by
acting on the IIB (IIA) theory with $i^{F_L}$ to give the $IIB^*$ ($IIA^*$) string theory.
This has the effect of multiplying all fields in the NS-R and R-R sectors  by $i$, so that on
rewrting in terms of real fields, the signs of  all terms in the action that are
bilinear in such fields, including the kinetic terms, are reversed, and exactly the same sign reversals 
arise from calculating the effective action in the formalism based on the imaginary spin operator $\hat S$, but with real
fields.
In either approach, 
  the RR field strengths $G_n $ of the type IIA (IIB) theory with $n$ even (odd) are related to the
field strengths $\hat G_{n\pm 1}$ of the T-dual type $IIB^*$ ($IIA^*$) theory by
$$G _{\mm_1 \dots \mm_n}=
\hat G_{0\mm_1 \dots \mm_n}, \qq 
G _{0\mm_1 \dots \mm_n}=-\hat G_{\mm_1 \dots \mm_n}
\ek
for any $\mm_i\ne 0$.

An important check on these signs and phases is given by the supergravity theories.
The T-duality between the IIA and IIB theories is reflected in the supergravity theories by the fact that the dimensional
reduction of   IIA supergravity to 8+1 dimensions is equivalent  to the reduction of IIB supergravity on a spacelike circle,
and the field redefiniton relating the two theories gives the T-duality rules, and in particular gives the RR field
transformations \tytty\ [\bergort], together with the transformations found by Buscher [\buscher].
Similarly, the IIA and $IIB^*$ supergravities have the same timelike reduction to 9+0 dimensions, as do the
IIB and $IIA^*$ supergravities, and the field redefinitions linking them give the T-duality transformations.

If there are fundamental strings whose world-sheets wrap around a compact time dimension, then there must be branes whose
world-volumes also wrap around compact time. For example, a fundamental string of the IIB string theory winding in time
is S-dual to a D-string winding time, and spatial T-duality leads to D$p$-branes of the IIB or IIA theories which also wrap
around the time dimension. 
The S-dual of a wrapped D5-brane gives a wrapped IIB NS 5-brane, and T-dualities relate these to wrapped NS-5-branes and KK
monopoles of the IIA and IIB theories.
It was argued in [\gibrap] that time-wrapping branes allows one to avoid  singularities and certain other undesirable features
in the brane
supergravity solutions.

A timelike T-duality takes a time-wrapped D$p$-brane (with $p+1$ dimensional Lorentzian world-volume) of the type IIA (IIB)
theory to an E$p$-brane of the type $IIB^*$ ($IIA^*$) string theory with a $p$-dimensional Euclidean world-volume [\hsta].
For example, a time-wrapped D-string is T-dual to an E$1$-brane of the $IIA^*$ theory whose world-line is spacelike and so
can be thought of as the world-line of a tachyon, but this tachyon is closely related to the tachyon discussed above, arising
from T-dualising a time-wrapped fundamental string.
The $IIB^*$ theory has fundamental strings with   Lorentzian world-sheets and E$2$-branes or \lq Euclidean strings' with
Euclidean 2-dimensional world-sheets, but the two are interchanged by $SL(2,\Z)$ S-duality transformations. This
duality with normal branes suggests that the tachyonic branes might not be as problematic as they at first appear.
As will be seen, there can be supersymmetric string theories in which the \lq fundamental strings'  are Euclidean, 
with Euclidean 
world-sheets embedded in a Lorentzian target space.
Similar considerations lead to branes of
various world-volume signatures occurring in the M-theories and strings of various signatures [\huku] and dualities link these,
so all should be on the same footing.

\chapter {T-Dual  of the $IIA_E$ Theory}

The $IIA_E$ string theory is the timelike reduction of M-theory, and is in 10 Euclidean dimensions. The fundamental strings
also have Euclidean world-sheets, as they arise from double timelike dimensional reduction of M2-branes.  In this section, we
will seek the  T-dual  of the
$IIA_E$ theory; it will be useful to denote it   as the
$IIB'$ theory. We will start by looking at the supergravity 
effective actions, and identify the perturbative branes as in [\ythe].
In particular, the dimensional reduction of the
$IIB'$ theory to 9 Euclidean dimensions should be the same as that of the  $IIA_E$ theory.

The bosonic action of the IIA supergravity is
$$\eqalign{
S_{IIA}= 
\int
 d^{10} x &
\sqrt{-g}\left[
 e^{-2 \Phi} \left( R+ 4(\partial   \Phi )^2  
-H^2 
\right)\right.
\cr &
\left. -G_2^2 - G_4^2 \right] + {4\over 3}\int G_4 \wedge G_4 \wedge B_2 + \dots
\cr}
\eqn\twoa$$
while that of IIB supergravity is
$$
\eqalign{S_{IIB}= 
\int
 d^{10} x &
\sqrt{-g}\left[
 e^{-2 \Phi} \left( R+ 4(\partial   \Phi )^2  
- H^2 
\right)\right.
\cr &
\left.-G_1 ^2 -  G_3 ^2 - G_5 ^2 \right] + \dots
\cr}
\eqn\twob$$
Here  $\Phi $ is the dilaton, $H=dB_2$ is the field strength of the NS-NS 2-form gauge field $B_2$ and 
$G_{n+1}=dC_n+\dots$ is the field strength for the RR $n$-form gauge field $C_n$. The field
equations derived from the IIB action \twob\ are supplemented with the self-duality constraint
$G_5=*G_5$. 

The corresponding actions for the type $II^*$ theories have RR kinetic 
terms with the opposite sign [\hsta]:
$$\eqalign{S_{IIA^*}= 
\int
 d^{10} x &
\sqrt{-g}\left[
 e^{-2 \Phi} \left(  R+ 4(\partial   \Phi )^2  
-H^2 
\right)\right.
\cr &
\left.
+G_2^2 +G_4^2 
\right]-{4\over 3}\int G_4 \wedge G_4 \wedge B_2 + \dots
\cr}
\eqn\twoas$$
and
$$
\eqalign{S_{IIB^*}= 
\int
 d^{10} x &
\sqrt{-g}\left[
 e^{-2 \Phi} \left( R+ 4(\partial   \Phi )^2  
- H^2 
\right)\right.
\cr &
\left. +G_1 ^2+  G_3 ^2+ G_5 ^2 \right] + \dots
\cr}
\eqn\twobs$$
The type $IIB^*$ theory has a scalar coset space
$${SL(2,\R)\over SO(1,1)}
\eqn\abc$$
and enjoys an $SL(2,\Z)$ U-duality, and again $G_5=*G_5$.

Reducing M-theory on a timelike $S^1$ gives
the $IIA_E$ string theory, and this has a supergravity limit given by the reduction of 11-dimensional
supergravity on a timelike circle. This is a theory in 10 Euclidean dimensions with bosonic action
$$\eqalign{S_{IIA_E}= 
\int
 d^{10} x &
\sqrt{g}\left[
 e^{-2 \Phi} \left(  R+ 4(\partial   \Phi )^2  
+ H^2 
\right)\right.
\cr &
\left.
+G_2^2 - G_4^2 
\right] +  {4\over 3}\int G_4 \wedge G_4 \wedge B_2+ \dots
\cr}
\eqn\twoae$$

 Reducing the type IIB or $IIA^*$ theory in the time direction gives the $IIB_9$ theory in 9 Euclidean dimensions   with
bosonic kinetic terms
$$\eqalign{S_{IIB_9}= 
\int
 d^{9} x &
\sqrt{g}\left[
 e^{-2 \Phi} \left( R+ 4(\partial   \Phi )^2  
-H^2  - d \chi ^2 + F_g^2 +F_B^2
\right)\right.
\cr & 
\left.-G_1^2 
+G_2^2 -G_3^2+ G_4^2 
\right]+\dots
\cr}
\eqn\ninb$$
and scalars taking values in 
$$
{SL(2,\R)\over SO(2)}\times \R^+
\eqn\slco$$
 while the timelike reduction of the IIA or $IIB^*$ theory gives the $IIA_{9}$ theory
$$\eqalign{S_{IIA_9}= 
\int
 d^{9} x &
\sqrt{g}\left[
 e^{-2 \Phi} \left(  R+ 4(\partial   \Phi )^2  
-H^2  - d \chi ^2 + F_g^2 -F_B^2
\right)\right.
\cr & \left.
+G_1^2 
-G_2^2 +G_3^2- G_4^2
\right]+\dots
\cr}
\eqn\nina$$
with scalars taking values in  
$$
{SL(2,\R)\over SO(1,1)}\times \R^+
\eqn\slcon$$
The fact that the timelike dimensional reduction of 
the IIB and $IIA^*$  (or IIA and $IIB^*$) theories gives the same 9-dimensional theory shows that timelike T-duality
should relate the two theories [\hsta].

On the other hand, reducing the $IIA_E$ supergravity to 9 dimensions gives the bosonic action
$$\eqalign{S_{IIA'_9}=
\int
 d^{9} x &
\sqrt{g}\left[
 e^{-2 \Phi} \left(  R+ 4(\partial   \Phi )^2  
+ H^2  - d \chi ^2+ F_g^2 +F_B^2
\right)\right.
\cr & \left.
+G_1^2 
+G_2^2 -G_3^2- G_4^2
\right]+\dots
\cr}
\eqn\nina$$
with scalars again taking values in  
\slcon.
This type $IIA'_9$ theory is related to the type $IIA_9$ theory by a field redefinition, which is   an $SL(2,\Z)$
transformation interchanging 
$B_2$ with $C_2$ and $C_1$ with $A^g$.
The dimensional reduction of 11-dimensional supergravity on space and then time gives the $IIA_9$ supergravity,
while reducing on time and then space gives the $IIA'_9$ supergravity. These two 9-dimensional supergravities are related by a
field redefinition and so are equivalent, so that reducing on time and then space gives the same result as reducing on space
and then time, as in [\CPS].
 The $IIA'_9$ theory can also be obtained by the timelike dimensional reduction of the following
9+1 dimensional bosonic action, which we will refer to as the $IIB'$ action:
$$
\eqalign{S_{IIB'}= 
\int
 d^{10} x &
\sqrt{-g}\left[
 e^{-2 \Phi} \left( R+ 4(\partial   \Phi )^2  
+ H^2 
\right)\right.
\cr &
\left. +G_1 ^2-  G_3 ^2+ G_5 ^2 \right] + \dots
\cr}
\eqn\twobp$$
This   differs from the $IIB^*$ action \twobs\ in   the signs of the kinetic terms of the 2-form gauge fields:
the NS-NS 2-form $B_2$ has a kinetic term of the right sign in the $IIB^*$ theory and the wrong sign in  the 
 $IIB'$ action, while the RR 2-form $C_2$ has a
kinetic term of the  wrong sign in the $IIB^*$ theory and the  right sign in  the 
 $IIB'$ action. 

The strong-coupling limit of the $IIB^*$ theory is given by acting with an $SL(2,\Z)$
transformation that takes $\Phi \to - \Phi$ (if $C_0=0$) and which interchanges $B_2$ and $C_2$ so
that, after a Weyl-rescaling of the metric, the $IIB^*$ action \twobs\ is replaced by the
$IIB'$ action \twobp, and the strong coupling limit of the   $IIB^*$ string theory is
the $IIB'$ string theory with field-theory limit \twobp.

This leads to the following scenario. 
The T-dual of the $IIA_E$ theory is a $IIB'$ string theory whose zero-slope limit is
the $IIB'$ supergravity \twobp, which is related to the $IIB^*$ supergravity by the field redefinition interchanging $B_2$
and $C_2$, so that the $IIB'$ and $IIB^*$ supergravities are equivalent field theories, with
two 2-form gauge fields which form an $SL(2)$ doublet and have kinetic terms of {\it opposite } signs.
 However, the  $IIB'$ and $IIB^*$
theories are {\it different } perturbative string theories. 
The 2-form with the right-sign kinetic term couples to strings with 1+1 dimensional world-sheets in the usual way
(the fundamental strings of the $IIB^*$ theory) while the one with the wrong-sign kinetic term couples to \lq Euclidean
strings' whose world-sheets are     spacelike surfaces with 2 Euclidean dimensions embedded in Lorentzian spacetime (the
E$2$-branes of the
$IIB^*$ theory). Thus   the $IIB^*$ and the $IIB'$ theories each 
have both Lorentzian and Euclidean strings, and which
of the various branes are the perturbative states can be seen using the arguments of [\ythe]. The
weakly coupled 
$IIB^*$ theory is a perturbative theory of the Lorentzian strings, with the Euclidean strings arising as non-perturbative
E-branes, while in the
$IIB'$ theory, the roles are reversed, and   the theory is a perturbative theory of the Euclidean strings
with the Lorentzian strings arising in the non-perturbative sector as D-strings on which the Euclidean strings can end.

Thus, at the non-perturbative level, there is a theory in 9+1 dimensions which has both  strings
with Lorentzian world-sheets and  strings
with Euclidean world-sheets, and whose zero-slope limit is the type $IIB^*$ supergravity with bosonic action
\twobs\ (or, equivalently, the type $IIB'$ supergravity \twobp).
The perturbative string theory based on the Lorentzian strings is the type $IIB^*$ string theory whose timelike T-dual is the 
type $IIA^*$ string theory, while the   perturbative string theory based on the Euclidean strings is the type $IIB'$ string
theory whose timelike T-dual is the  type $IIA_E$ string theory, which is again a theory of Euclidean world-sheets. 
The $IIB^*$ and  $IIB'$ string theories are S-dual, so one is the strong-coupling limit of the other.
We see that the T-dual of the  $IIA_E$ string theory is  not a version of the IIB theory with Euclidean target space, but a
version with Euclidean world-sheet instead. We also learn that the diagram in figure 3 is essentially correct, so long as the 
$IIB^*$ theory occurring at $R=T=0$ is taken as the full non-perturbative theory; the coupling constant is given by the ratio
$R/T$, as in [\asp], and so taking $R$ to zero and then $T$ to zero gives the $IIB^*$ string theory
while reversing the order gives its strong coupling limit, the $IIB'$ string.

The key feature here is that the 2-form with the wrong sign kinetic term couples to strings with Euclidean world-sheets, and
a Euclidean string on a spacelike circle of radius $R$ is T-dual to a Euclidean string theory
on a timelike circle of radius $1/R$, so that 
 whereas T-duality of Lorentzian strings does not change the
space-time signature,   T-duality of Euclidean ones does. 
 In the next section, we will show why this is the case; 
in later sections, this will lead to further circumstances in which
dualities can change  space-time signature.

\chapter{Euclidean Strings and T-Duality}

Consider a string theory  with a flat target space with a circular dimension with coordinate $X\sim X+1$, so that
the world-sheet action includes  
$$S=\ll \2 \int d^2 \ss \pa _ a X \pa ^aX + \dots
\eqn\abcfdfd$$
where $\ll=R^2 $ for a space-like circle of radius $R$ and $\ll=-R^2  $ for a timelike circle of radius $R$, and
$\ss^a$ are the world-sheet coordinates. T-duality can be studied by gauging the shift symmetry $X \to X+ c$ 
by coupling to a world-sheet gauge field $A_a$ and adding a
term with a Lagrange multiplier field $Y$ imposing the constraint that the gauge field is flat
[\buscher-\TD]. This gives the action
$$S=  \int d^2 \ss\left( \2 \ll D _ a X D ^aX + \ee ^{ab}A_a \pa _b Y+\dots \right)
\eqn\abc$$
where
$$D_aX=\pa _a X-A_a
\eqn\abc$$
Then the Lagrange multiplier imposes the constraint $F_{ab}=0$ $(   F_{ab}= \pa _a A_b-\pa_b A_a)$ and implies $A$ is  pure
gauge and can be gauged away, recovering  the action \abcfdfd.  The multiplier field $Y$ takes
values in a circle, $Y \sim Y+1$, so that the winding modes in $Y$ are responsible for  the
elimination of flat connections with non-trivial holonomy.  On the other hand, the terms
involving $X$ can be gauged away, after which
$A_a$ is an auxiliary field which can be eliminated to give
$$S= \ll ' \2 \int d^2 \ss  \pa _ a Y \pa ^aY + \dots
\eqn\abc$$
where 
$$
\ll '=-{\eee \over \ll}
\eqn\abc$$
and $\eee$ is the coefficient in the identity
$$
\ee^{ab}\ee ^{cd}= \eee \left( h^{ac}h^{bd}-h^{ad}h^{bc}\right)
\eqn\epid$$
where $h_{ab}$ is the world-sheet metric. Then $\eee=-1$ for a Lorentzian world-sheet and $\eee=+1$ for a Euclidean
world-sheet. Thus for a Lorentzian world-sheet, $\ll ' = 1/\ll$ and a spacelike (timelike) circle of radius $R$ 
in which $X$ takes values has been exchanged for the dual spacelike (timelike) circle of radius $1/R$ 
in which $Y$ takes values, so the signature is unchanged.
On the other hand, for a Euclidean world-sheet, $\ll ' = -1/\ll$ and a spacelike (timelike) circle of radius $R$ 
in which $X$ takes values has been exchanged for the dual timelike  (spacelike) circle of radius $1/R$ 
in which $Y$ takes values, so the signature of the $X$ coordinate is reversed.
The T-duality corresponds to the map $X\to Y$ where $\pa_a X = \ee _a{}^b \pa _b Y$, so that
$(\partial X)^2=-\eee (\partial Y)^2$.

More generally, consider a non-linear sigma-model
$$S= \2   \int d^2 \ss \sqrt{\vert h \vert }\left(g_{mn} \pa _ a X^m \pa ^aX ^n +B_{mn} \ee^{ab} \pa _aX^m \pa _b X^n + \Phi R
\right)
\eqn\sigm$$
representing a string moving in  a target space with coordinates $X^m$, metric $g_{mn}$ and background 2-form gauge field
$B_{mn}$ and dilaton $\Phi$. Suppose the target space has an isometry generated by a Killing vector $k^m$ and that the Lie
derivatives of $H_{mnp}$ and the dilaton vanish. Then there is a symmetry $\dd X= c k^m$ for constant parameter $c$ and this
can be gauged if the global obstruction of [\HS] is absent. In that case,   a Lagrange multiplier
term can again be added to the gauged action to obtain [\buscher-\TD]
$$S=S_{gauged }[X^m, A_a]  +  \int d^2 \ss   \ee ^{ab}A_a \pa _b Y 
\eqn\abc$$
where $S_{gauged }$ is the gauged action given in [\HS].
If the isometry orbits are compact, the Lagrange multiplier can again be eliminated to regain the action
\sigm, or one can integrate over the gauge fields to obtain the T-dual sigma-model with background
fields $\ti g_{mn},\ti B_{mn},\ti \Phi$. Choosing adapted coordinates $X^m=(X^0, X^\aa)$ where
$k=\pa/\pa X^0$ so that the isometry generates shifts in
$X^0$,\foot{We use the notation of [\buscher-\TD], but do not specify here whether $X^0$ is a
spacelike or timelike coordinate.} the dual background fields are
$$\eqalign{\ti g_{00} &= -\eee{1 \over g_{00}} \cr
\ti g_{0\aa} &= -\eee {B_{0\aa} \over g_{00}} \cr
\ti B_{0\aa} &= {g_{0\aa} \over g_{00} }\cr
\ti g_{ \aa\bb} &=  g_{ \aa\bb}- {g_{0\aa}g_{0\bb}-(-\eee)B_{0\aa} B_{0\bb}\over g_{00} }\cr
\ti B_{ \aa\bb} &=  B_{ \aa\bb}- {g_{0\aa}B_{0\bb}-B_{0\aa} g_{0\bb}\over g_{00} }\cr
\ti \Phi=\Phi - \2 log \vert g_{00}\vert \cr
}
\eqn\abc$$
 These are the same as the transformations of   [\buscher], except that   the factors of $
-\eee$, with $\eee= -1 $ for Minkowski world-sheets and $\eee=1$ for Euclidean world-sheets, 
changes some of the signs for Euclidean strings.  In particular, as $\ti g_{00}  = -\eee{/ g_{00}}$,
the target space signature changes if $\eee=1$.

The metric beta-function for the sigma-model \sigm\ is 
$$\bb ^g_{mn}=R_{mn} - (-\eee) H_{mpq}H_n{}^{pq} +\dots
\eqn\abc$$
as the identity \epid\ is used in the Feynman graphs leading to the $H^2$ term.
As a result, the effective action generating the conditions for sigma-model conformal invariance is
$$ S= \int d^DX \sqrt{\vert g \vert }e^{-2\Phi}\left(R+4 \nabla \Phi ^2 - (-\eee) H^2 
\right)+\dots
\eqn\abc$$
and the sign of the $B_2$ kinetic term is indeed reversed for Euclidean world-sheets ($\eee=1$), as expected from the last
section.

Note that the Euclidean strings considered here should be distinguished from the Wick-rotated strings used in Euclidean path
integrals.
Consider for example the sigma-model \sigm\ with Lorentzian world-sheet and Lorentzian target space. In the Euclidean path
integral quantization, the target space is (usually) Wick-rotated or analytically continued to a Euclidean space
and the path integral is taken to be over all Euclidean world-sheet metrics. The functional integral is weighted with 
$e^{-S_E}$ where $S_E$ is the Euclideanised action which in particular has an {\it imaginary }
Wess-Zumino term:
$$S_E= \2   \int d^2 \ss \sqrt{\vert h \vert }\left(g_{mn} \pa _ a X^m \pa ^aX ^n +iB_{mn} \ee^{ab} \pa _aX^m \pa _b X^n + \Phi
R
\right)
\eqn\abc$$
The factor of $i$ in front of the  Wess-Zumino term is needed to make 
the functional integral well-defined for topologically
non-trivial $B_2$ fields   and gives an extra sign to cancel the one coming from the identity \epid\ on
changing from Lorentzian to Euclidean world-sheets.  The resulting target space effective action  is
then 
$$ S= \int d^DX \sqrt{\vert g \vert }e^{-2\Phi}\left(R+4 \nabla \Phi ^2 -   H^2 
\right)+\dots
\eqn\abc$$
with the correct signs, and the T-duality transformations are those of Buscher, without any sign-changes.
The action for the Euclidean strings that was considered above has a {\it real} Wess-Zumino term and
so has some   signs changed in the T-duality rules and the effective action. Formally, one can think
of   using the   action \sigm\  in a
  path-integral weighted with $e^{iS}$, and then consider 
various Wick rotations or analytic continuations as appropriate to
 define the functional integral; these include Wick rotations of timelike target space coordinates
$X$ as well as the world-sheet time.

\chapter{More Type II String Theories}

The type $IIB'$ theory has Euclidean fundamental strings and so,  by the arguments of the last section, T-duality changes the signature.  Compactifying on a timelike circle and T-dualising gives the $IIA_E$ theory in 10 Euclidean dimensions, while using a spacelike circle
will give a theory in 8+2 dimensions. 
The supergravity action for the theory in 8+2 dimensions is 
determined by requiring its timelike reduction to 8+1 dimensions be the same as the spacelike reduction of the $IIB'_{9+1}$
supergravity, and is again of the type IIA form but with certain sign changes. The theory in 8+2 dimensions will be referred
to as the type $IIA_{8+2}$ theory. Both the $IIA_E$ and $IIA_{8+2}$ theories have  a $B_2$ field with the wrong sign kinetic
term and so both have Euclidean fundamental strings. 

In the same way, a spacelike T-duality of the $IIA_{8+2}$ theory will give a
$IIB_{7+3}$ theory in 7+3 dimensions which has a type IIB-style action but with some sign changes (given in table 1 of
section 12).
 It has scalars in the coset space $SL(2,\R)/SO(2)$,
and both the 2-forms $C_2$ and $B_2$ have kinetic terms of the wrong sign, so that there are $(p,q)$
strings with Euclidean world-sheets. A spacelike T-duality then takes this to a $IIA_{6+4}$ theory
in 6 spatial and 4 time
dimensions, and a further spacelike T-duality takes this to a IIB-like theory in 5+5 dimensions, which will be denoted   the  
$IIB'_{5+5}$ string theory, as it is similar to the
$IIB'_{9+1}$ theory. It has fundamental Euclidean strings and Lorentzian D-strings, and the strong coupling limit is a type 
$IIB^*_{5+5}$ string theory with   perturbative Lorentzian strings
and E2-branes (or Euclidean D-strings).  
This has a $IIA^*_{5+5}$ theory as a spacelike T-dual, and a $IIA _{5+5}$ theory as a timelike
T-dual, and finally the $IIA _{5+5}$ has a $IIB _{5+5}$ theory as a spacelike T-dual, and this
latter is the timelike T-dual of the
 $IIA^*_{5+5}$ theory. 
The T-dualities relating the theories in 5+5 dimensions are shown in figure 4,
and is very similar to the 9+1 dimensional case depicted in   figure 1.

\vskip .5 cm

{\vbox{

%%Begin InstantTeX Picture
\let\picnaturalsize=N
\def\picsize{3.0in}
\def\picfilename{SigFigs/fig13.eps}
%If you do not have the picture file add:
%\let\nopictures=Y
%to the beginning of the file.
\ifx\nopictures Y\else{\ifx\epsfloaded Y\else\input epsf \fi
\let\epsfloaded=Y
\centerline{\ifx\picnaturalsize N\epsfxsize \picsize\fi \epsfbox{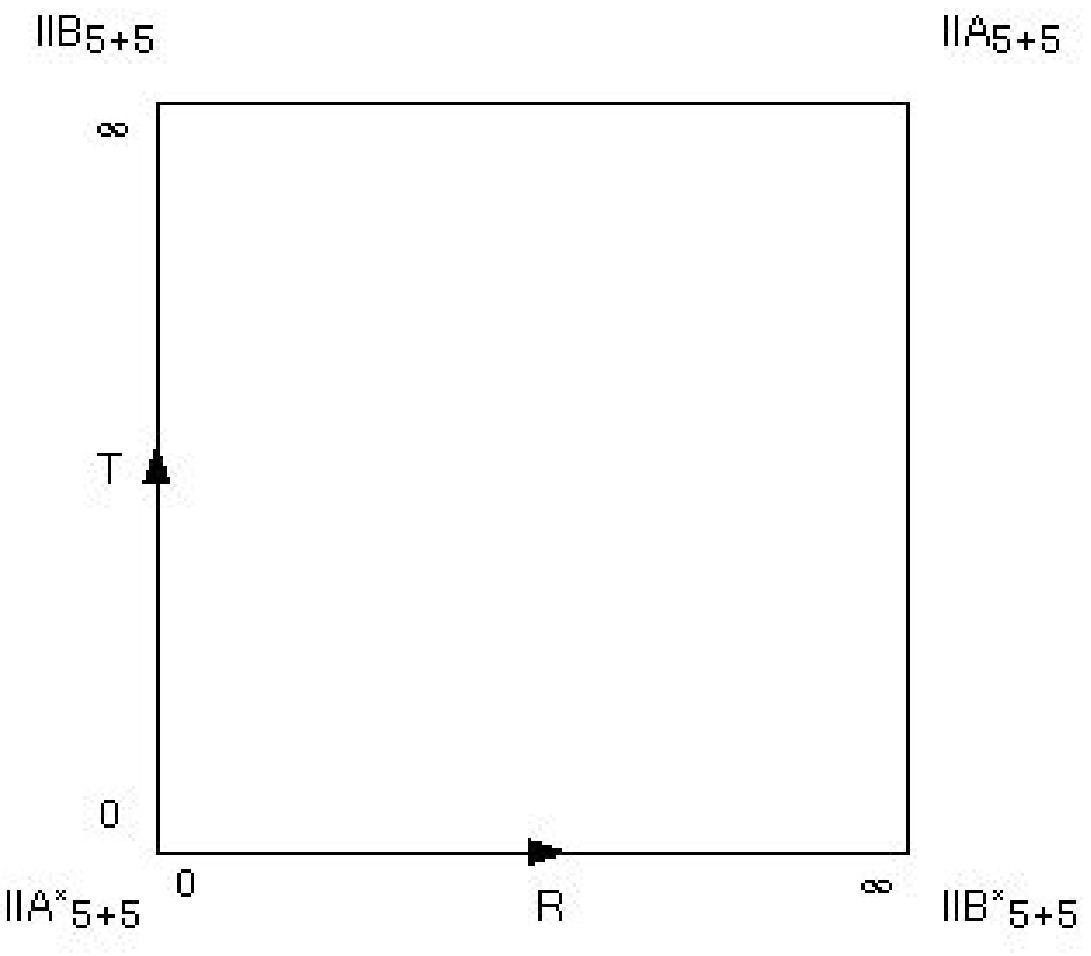}}}\fi
%%End InstantTeX Picture

{\bf Figure 4}  The moduli space for the type $IIA_{5+5}$ theory compactified on a Lorentzian torus with 
timelike radius $T$ and spacelike radius $R$. 
}}

\vskip .5 cm

\noindent Spacelike T-duality relates the IIA and IIB theories and the $IIA^*$ and $IIB^*$
theories, while  timelike T-duality relates
the
$IIA^*$ and IIB theories and the IIA  and $IIB^*$
theories. The type $IIB^*_{5+5} $  string and the  type $IIB'_{5+5} $  string are different perturbative limits of the same
non-perturbative theory. After this, further T-dualities give a set of theories similar to 
those just listed but with space and
time interchanged.
Thus there are $IIA,IIB,IIA^*IIB^*$ theories with signature 1+9 (one space and 9 time) together
with theories in 0+10, 2+8, 3+7 and 4+6 dimensions. 
The  supergravity theory in  signature  $(s,t)$   and its mirror with signature
$(t,s)$  are equivalent, as will be discussed in section 11.
The $IIA_{5+5}$ and the $IIA^*_{5+5}$ actions are equivalent, related by taking $g_{\mu \nu } \to
-g_{\mu \nu }$, and changing the overall sign of the action, while such a mirror transformation
takes the $IIB_{5+5}$ theory to itself, and the $IIB_{5+5}^*$ theory to itself.
 A   map through
this maze of T-dual 10-dimensional theories is given in figure 5.
Note that this includes the square of theories in 9+1 dimensions given in figure 1, 
its mirror in 1+9 dimensions and the square of theories in 5+5 dimensions given in figure 4.
Reflection about the central horizontal line 
on which the
$IIB_{5+5}$ theories lie 
 takes each theory to its mirror.

\vskip .5 cm

{\vbox{

%%Begin InstantTeX Picture
\let\picnaturalsize=N
\def\picsize{5.0in}
\def\picfilename{SigFigs/map.eps}
%If you do not have the picture file add:
%\let\nopictures=Y
%to the beginning of the file.
\ifx\nopictures Y\else{\ifx\epsfloaded Y\else\input epsf \fi
\let\epsfloaded=Y
\centerline{\ifx\picnaturalsize N\epsfxsize \picsize\fi \epsfbox{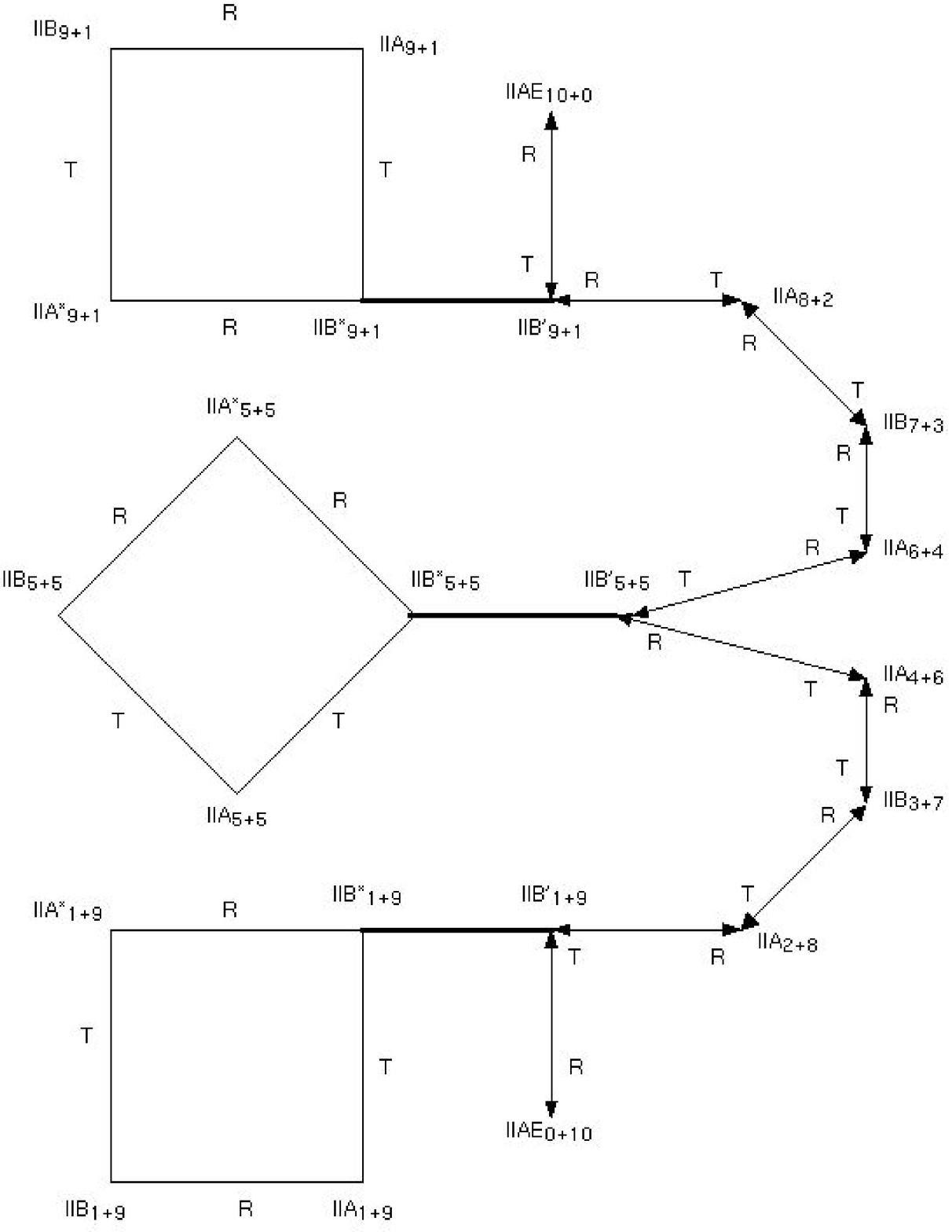}}}\fi
%%End InstantTeX Picture

{\bf Figure 5}  The dualities linking the type II string theories in 10 dimensions. 
A bold line represents an S-duality interchanging strong and weak coupling, 
a line with an $R$ indicates a spatial T-duality, a line with a $T$ indicates a timelike T-duality, and a double-ended arrow 
with both an $R$ and a $T$ indicates a signature changing T-duality, where the theory at the $R$ end compactified on a
spacelike circle of radius $R$ is T-dual to the theory at the $T$ end on a timelike circle of radius $T=1/R$.}}

The $IIA _{5+5}$ and $IIB _{5+5}$  theories have actions in which all of the kinetic terms have the
\lq right sign', so that the corresponding actions are the same as those for the usual $IIA$ and
$IIB$ theories
\twoa,\twob, but with different signature. In 5+5 dimensions there are Majorana-Weyl spinors as well
as real self-dual 5-forms, just as in 9+1 dimensions. The $IIA _{5+5}^*$ and $IIB _{5+5}^*$ string theories
are obtained from the $IIA _{5+5}$ and $IIB _{5+5}$ string  theories in the same way as the corresponding theories in 9+1
dimensions, by acting with $i^{F_L}$, so that   the signs of the RR gauge field kinetic terms are reversed so that the
supergravity actions are of the form
\twoas,\twobs. The
$IIA _{5+5}$ and $IIB _{5+5}$  theories have untwisted superalgebras
$$ \{Q_i,Q_j \} =  (\ggg ^\mm C^{-1}) P_\mm \dd _{ij}
\eqn\untw$$ while the $IIA _{5+5}^*$ and $IIB _{5+5}^*$ theories have the twisted superalgebra
\salg.
The IIA and IIB string theories in signatures 9+1,5+5,
and the $IIA^*$ and $IIB^*$ string theories in signature 1+9 
 are the only string theories in which all   signs and phases are
the conventional ones.

\chapter{The Strong-Coupling Limit of the Type $IIA^*$ Theory}

In this section, we will investigate the strong-coupling limit of the type $IIA^*$ theory.
For the IIA theory, there are D0-branes with mass $m\sim n/g$ for all integers $n$
where $g$ is the
string coupling, and these  become light at strong coupling and are identified with a Kaluza-Klein 
tower of states from the decompactification of an extra spatial dimension [\wittz].
For the $IIA^*$ theory, there are E1-branes with tachyonic mass  $m$, $m ^2=-\mm^2$ with $\mm$ real and 
$\mm \sim n/g$ for all integers $n$. Then $\mm \to 0$ at strong coupling and there is again an infinite tower of states 
that become massless in the limit and which fit into supergravity multiplets.
Whereas the compactification of a spatial dimension gives a Kaluza-Klein 
tower of massive states, the compactification of a timelike dimension gives a Kaluza-Klein 
tower of tachyonic states, and in either case the tower becomes light as the extra dimension
decompactifies.
Thus it is natural to identify the strong coupling limit of the $IIA^*$ theory as
one in which an extra {\it timelike} dimension decompactifies to give a theory in 9+2 dimensions.

This result can be checked in a number of ways.
It is certainly to be expected that   the strong
coupling limit should lead to the decompactification of an 11'th dimension, and the field-theory
limit of the 11-dimensional theory should, on dimensional reduction, give the
$IIA^*$ supergravity \twoas\ in 9+1 dimensions.
This 11-dimensional field theory cannot be the  usual 11-dimensional supergravity [\CJ]
with bosonic action
$$%\eqalign{
S_{M}= 
 \int
 d^{11} x 
%&
\sqrt{-g}\left[
   R - G_4^2 \right] + {4\over 3}\int G_4 \wedge G_4 \wedge C_3  
%\cr}
\eqn\mmmm$$
since the  spatial reduction  of this gives the
$IIA$ theory, not the $IIA^*$  theory, while the timelike reduction gives a Euclidean theory, not one in 9+1 dimensions.
As the $IIA^*$  theory has a kinetic term for $C_3$ of the wrong sign, a 
  natural guess would be a 10+1 dimensional field theory 
of the form \mmmm\ but with the sign of the kinetic term of $G_4$ reversed.
However, the spatial reduction of such an action would   give kinetic terms for $B_2$ and $C_1$ differing in sign from those of
the
$IIA^*$ theory \twoas. However, the fact that the kinetic term of $C_1$ in \twoas\ has the wrong
sign means that if it is to be interpreted as the graviphoton from the reduction of an
11-dimensional metric, then the reduction must be on a timelike circle, not a space-like one.
Consider, then, the following action in 9+2 dimensions:
$$ \eqalign{
S_{M}= 
 \int
 d^{11} x 
 &
\sqrt{g}\left[
   R + G_4^2 \right] + {4\over 3}\int G_4 \wedge G_4 \wedge C_3  
 \cr}
\eqn\mmms$$
Its dimensional reduction on  a timelike circle indeed gives the $IIA^*$  
 action, 
and if there is any 11-dimensional supergravity theory whose dimensional reduction gives the $IIA^*$ theory, it must have this
bosonic sector.
The strong
coupling limit of the $IIA^*$ theory is then a theory in 9+2 
dimensions which we will denote as $M^*$ theory. The effective
supergravity theory is given by coupling \mmms\ to a 
 gravitino which, as will be seen in section  11, is a pseudo-Majorana spinor (with 32 real components).

 Another way of obtaining the same theory is as follows. Consider first M-theory
compactified on a space-like 3-torus, $T^3$. In the limit that the torus shrinks to zero size, one
recovers M-theory [\fhrs]; dimensional reduction   gives a theory in 7+1 dimensions, but
the modes from membranes wrapping around each of the three 2-cycles become light as the cycles
shrink, and these light modes signal the opening up of three new spatial dimensions. If instead  
M-theory is compactified on a Lorentzian 3-torus, $T^{2,1}$, then in the limit that the torus
shrinks to zero size,  one again expects 3 new dimensions to open up, in addition to the 8
Euclidean dimensions  from the dimensional reduction. 
However, in this case there is one     
Euclidean $T^2$ 2-cycle and two Lorentzian $T^{1,1}$ 2-cycles, and
we shall see that the shrinking Euclidean 2-cycle gives an extra space dimension while each
shrinking Lorentzian 2-cycle gives an extra time dimension, resulting in a theory in 9+2 dimensions.

 M-theory on a Euclidean $T^2$  
shrinking  to zero size gives  8+1 dimensions, as in the field theory limit, together with an extra
spatial dimension from membranes wrapped on $T^2$, to give the IIB string in 9+1 dimensions. On the
other hand, as we have seen,  M-theory on a Lorentzian  $T^{1,1}$   shrinking  to zero size gives  9
Euclidean  dimensions, as in the field theory limit, together with an extra timelike dimension from
membranes wrapped on $T^{1,1}$, to give the
$IIB^*$ theory in 9+1 dimensions. Thus membranes wrapped on a shrinking $T^{2,0}$ give an extra 
space dimension while
those wrapped on a shrinking $T^{1,1}$ give an extra time dimension.
This also follows from the fact that T-duality of Euclidean strings changes the signature.
Then for M-theory on a shrinking $T^{2,1}$, one spatial and two timelike dimensions
 open  up, to give $M^*$ theory in 9+2
dimensions.
  M-theory and $M^*$ theory are then linked by a duality, and so are not independent theories: M-theory on a Lorentzian torus
$T^{2,1}$ is equivalent to $M^*$-theory compactified on a dual torus which, as we will see in section 10, is a torus $T^{1,2}$
with the opposite signature.

\chapter{ Compactifications of $M^*$ Theory}

Consider now reductions of $M^*$ theory, analogous to the ones considered above for   M-theory. Reduction on a timelike
circle gives the $IIA^*$ theory in 9+1 dimensions, with coupling constant related to the radius so that the strong coupling
limit of the 
$IIA^*$ theory is a decompactification to $M^*$ theory in 9+2 dimensions.
On the other hand, reduction on a spacelike circle gives a string theory in 8+2 dimensions, which is the
$IIA_{8+2}$ string theory obtained in   section 6. Considering toroidal reductions of $M^*$ theory clarifies some of the
dualities represented in figure 5.

First, reduction of $M^*$ theory on a torus $T^{0,2}$ with 2 timelike circles gives the moduli space in figure 6.

\vskip .5 cm

{\vbox{
%%Begin InstantTeX Picture
\let\picnaturalsize=N
\def\picsize{3.0in}
\def\picfilename{SigFigs/fig3.eps}
%If you do not have the picture file add:
%\let\nopictures=Y
%to the beginning of the file.
\ifx\nopictures Y\else{\ifx\epsfloaded Y\else\input epsf \fi
\let\epsfloaded=Y
\centerline{\ifx\picnaturalsize N\epsfxsize \picsize\fi \epsfbox{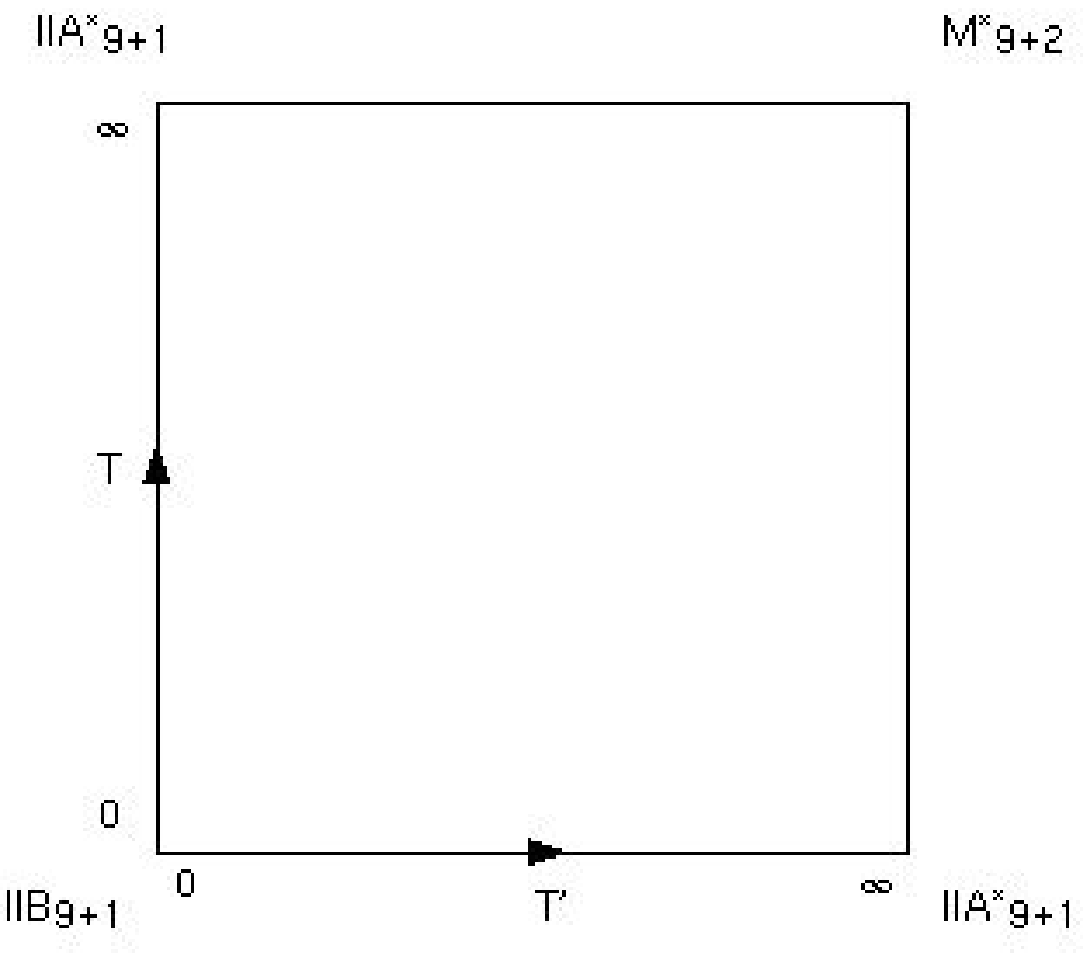}}}\fi
%%End InstantTeX Picture

{\bf Figure 6}  The moduli space for the compactification of $M^*$-theory from $9+2$ dimensions on a torus $T^{0,2}$ with 
two timelike circles of radii $T,T'$.
The figure should be identified under reflection in the $T=T'$ diagonal.
}}

\vskip .5 cm

\noindent The $IIA^*$ and IIB theories are related by timelike T-duality and the limit as the $T^{0,2}$ shrinks to zero size
gives the IIB theory. 

Reduction of  $M^*$ theory on a Lorentzian torus $T^{1,1}$ gives the following diagram:
\vskip .5 cm

{\vbox{

%%Begin InstantTeX Picture
\let\picnaturalsize=N
\def\picsize{3.0in}
\def\picfilename{SigFigs/fig4.eps}
%If you do not have the picture file add:
%\let\nopictures=Y
%to the beginning of the file.
\ifx\nopictures Y\else{\ifx\epsfloaded Y\else\input epsf \fi
\let\epsfloaded=Y
\centerline{\ifx\picnaturalsize N\epsfxsize \picsize\fi \epsfbox{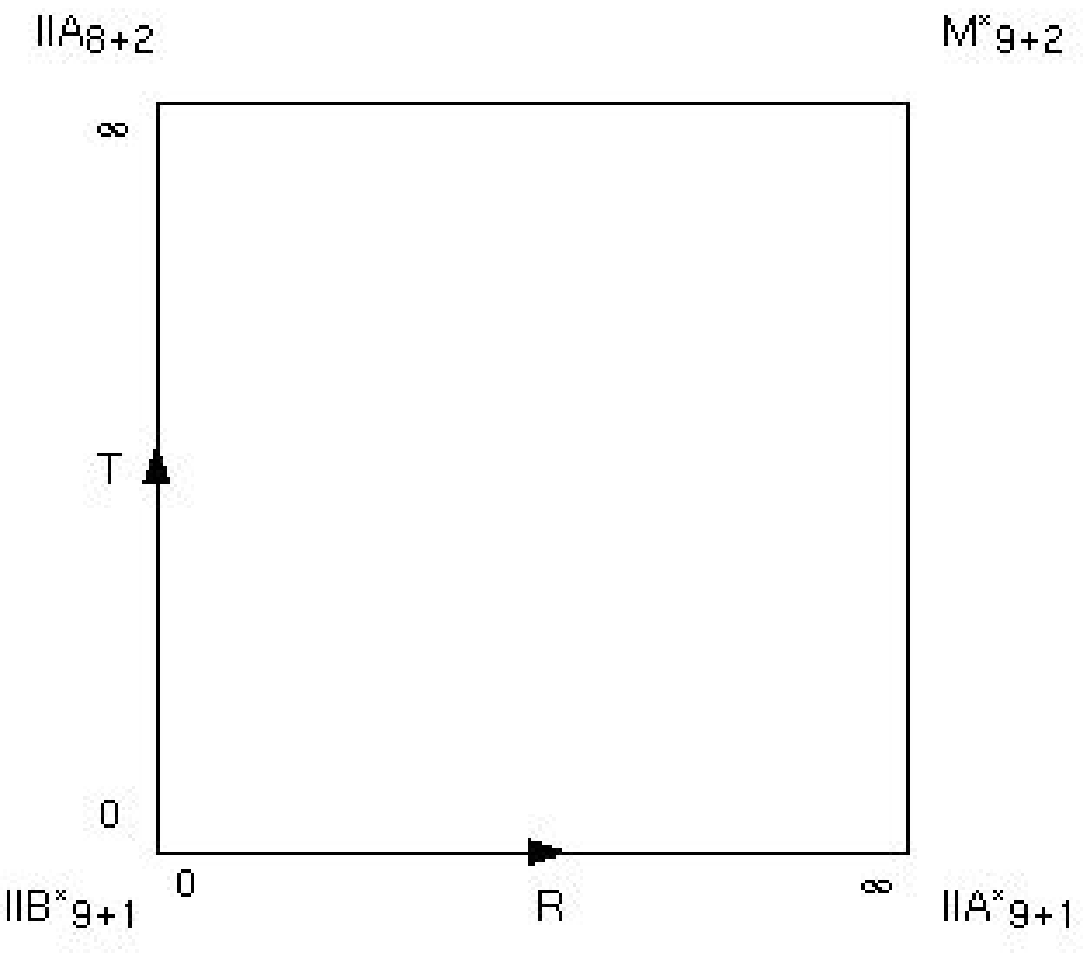}}}\fi
%%End InstantTeX Picture

{\bf Figure 7}  The moduli space for the compactification of $M^*$-theory from 
$9+2$ dimensions on a  Lorentzian torus $T^{1,1}$  with  spacelike radius $R$ and  timelike radius  $T$.}}

\vskip .5 cm

\noindent Compactification first on time leads to the T-duality between the $IIA^*$ theory on a spatial circle
of radius $R$ and the 
$IIB^*$ theory on a spatial circle of radius $1/R$.
  From
section 4, the
$IIB'$ string theory is the strong coupling limit of the $IIB^*$ string theory, and 
the $IIB'$ and  $IIA_{8+2}$ theories both have Euclidean fundamental strings.
There is a signature-changing T-duality between the $IIA_{8+2}$ theory on a timelike circle of radius $T$ and the 
$IIB'$ string theory on a spatial circle of radius $R=1/T$, and this arises in fig. 5 from
 compactifying first on space.
In the limit in which the $T^{1,1}$ 
shrinks to zero size, we obtain the $IIB^*$ theory with coupling constant $g \sim R/T$.
The strong coupling  limit of the $IIA_{8+2}$ theory is the $M^*$ theory, with a spacelike circle decompactifying.

Compactifying the $IIA_{8+2}$ theory on a spacelike circle of radius $R $ is equivalent to compactifying a T-dual theory on a
timelike circle
 of radius $1/R $, and the T-dual theory must be in signature 7+3, and we will denote it as $IIB_{7+3}$. 

The moduli space for reduction of $M^*$ theory on a spatial torus $T^{2,0}$ is shown in figure 8.
The $M^*$ theory picture leads to the signature-changing T-duality between the
$IIB_{7+3}$ theory and the $IIA_{8+2}$ theory, and
the limit in which
the $T^{2,0}$ shrinks to zero size gives the $IIB_{7+3}$ theory.

\vskip .5 cm

{\vbox{

%%Begin InstantTeX Picture
\let\picnaturalsize=N
\def\picsize{3.0in}
\def\picfilename{SigFigs/fig5.eps}
%If you do not have the picture file add:
%\let\nopictures=Y
%to the beginning of the file.
\ifx\nopictures Y\else{\ifx\epsfloaded Y\else\input epsf \fi
\let\epsfloaded=Y
\centerline{\ifx\picnaturalsize N\epsfxsize \picsize\fi \epsfbox{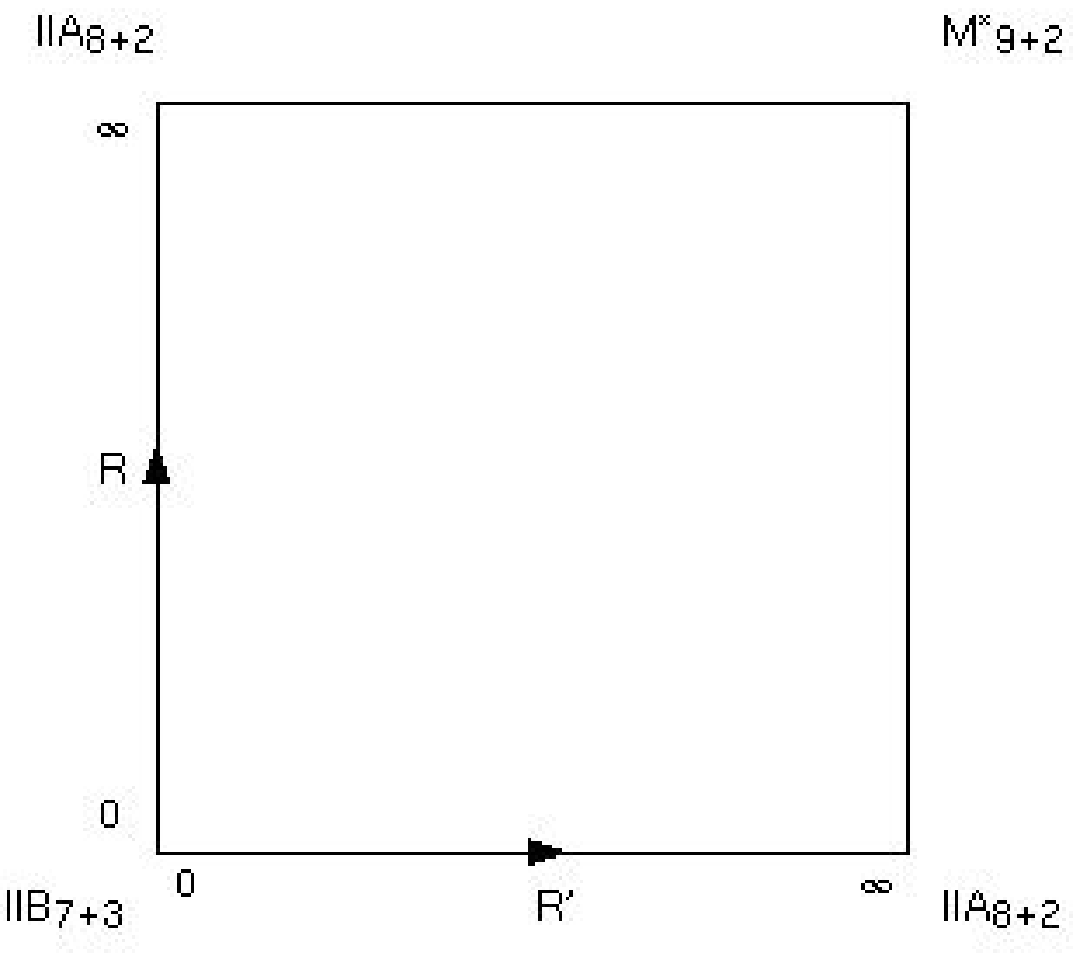}}}\fi
%%End InstantTeX Picture

{\bf Figure 8}  The moduli space for the compactification of $M^*$-theory from $9+2$ dimensions on a
 Euclidean torus $T^{2,0}$ with two spacelike circles of radii $R,R'$.
The figure should be identified under reflection in the $R+R'$ diagonal.
}}

\vskip .5 cm

\chapter{More Strong Coupling Limits}

The $IIA_{5+5}$ string in 5+5 dimensions is very similar to the usual
$IIA$ string in 9+1 dimensions.
In particular, it has D0-branes that become light at strong coupling,
signalling the decompactification of a new spacelike dimension, so that its strong coupling limit
is a theory in signature $6+5$ dimensions which we will refer to as the  $M'_{6+5}$ theory.
This is also the   strong coupling limit of the $IIA_{6+4}$ theory, which has E1-branes that become light at strong couling,
signalling an extra time dimension.
 A  supergravity theory
in 6+5 dimensions with lagrangian $R-(dC_3)^2+...$ is the unique 11 dimensional theory that can be dimensionally reduced to
give the $IIA_{5+5}$ supergravity (by reducing on a spacelike circle) or the $IIA_{6+4}$ supergravity (by reducing on a
timelike circle).
Similarly, the strong-coupling limit  of the
$IIA_{5+5}^*$ string and the $IIA_{4+6}$ string is a reversed-signature $M'_{5+6}$ theory in 5+6
dimensions. 
The   strong coupling limits 
of the $IIA_{2+8}$ string and the $IIA_{1+9}^*$ strings are
a reversed signature $M^*_{2+9} $ theory in   2+9 dimensions
and 
the   strong coupling limits 
of the $IIA_{0+10}$ string and the $IIA_{1+9}$ strings are
a reversed signature $M_{1+10} $ theory in   1+10 dimensions.

Consider now the compactification of the $M'$ theory on circles and 2-tori of various signatures. 
The reduction on a timelike circle gives the $IIA_{6+4}$ theory, while on a spacelike circle gives a $IIA_{5+5}$ theory 
in 5+5 dimensions, which is T-dual to a $IIB_{5+5}$ theory 
in 5+5 dimensions on a spacelike circle.
The moduli spaces for the compactifications on the tori $T^{2,0}$, $T^{1,1} $ and $T^{0,2}$ are shown in   figures 9,10 and 11.

\vskip .5 cm

{\vbox{

%%Begin InstantTeX Picture
\let\picnaturalsize=N
\def\picsize{3.0in}
\def\picfilename{SigFigs/fig6.eps}
%If you do not have the picture file add:
%\let\nopictures=Y
%to the beginning of the file.
\ifx\nopictures Y\else{\ifx\epsfloaded Y\else\input epsf \fi
\let\epsfloaded=Y
\centerline{\ifx\picnaturalsize N\epsfxsize \picsize\fi \epsfbox{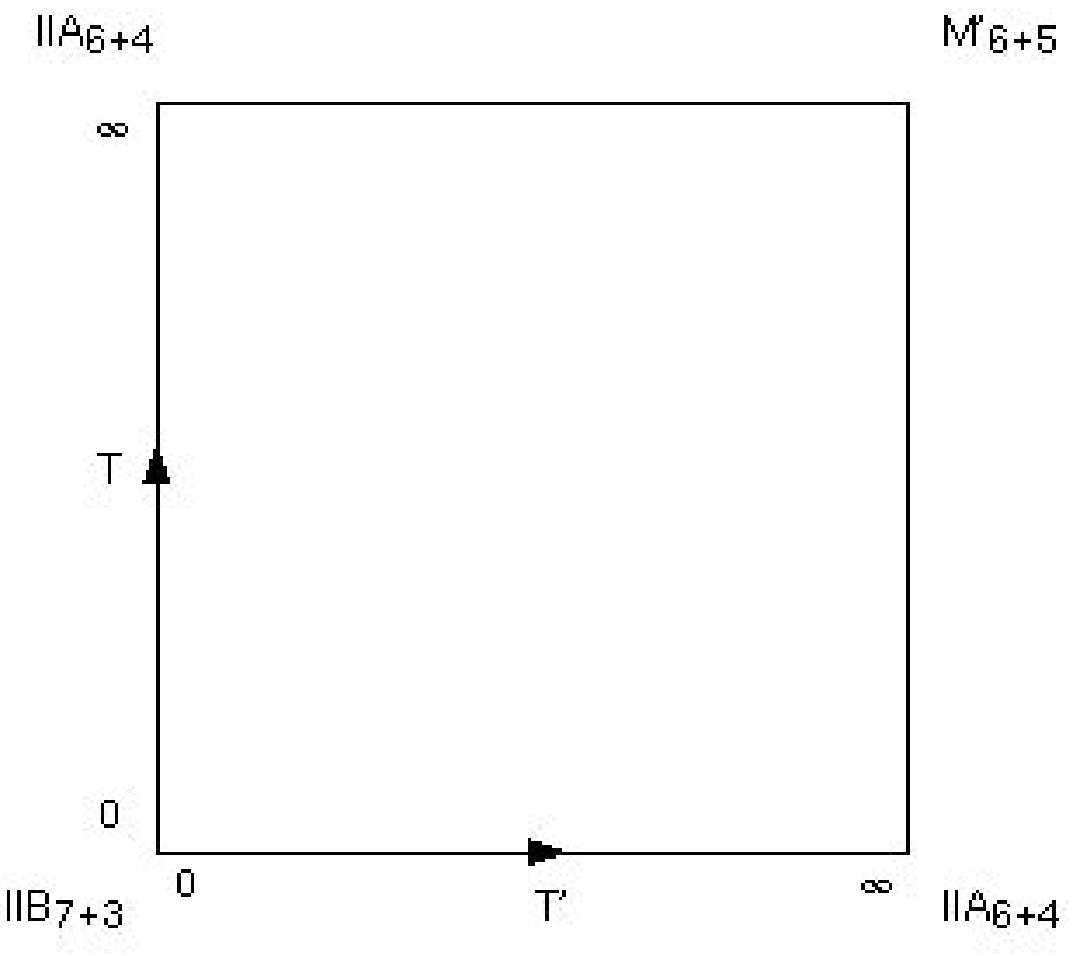}}}\fi
%%End InstantTeX Picture

{\bf Figure 9}  The moduli space for the compactification of $M'$-theory from $6+5$ dimensions on a 
torus $T^{0,2}$ with two timelike circles of radii $T,T'$.
The figure should be identified under reflection in the $T=T'$ diagonal.
}}

\vskip .5 cm

{\vbox{

%%Begin InstantTeX Picture
\let\picnaturalsize=N
\def\picsize{3.0in}
\def\picfilename{SigFigs/fig7.eps}
%If you do not have the picture file add:
%\let\nopictures=Y
%to the beginning of the file.
\ifx\nopictures Y\else{\ifx\epsfloaded Y\else\input epsf \fi
\let\epsfloaded=Y
\centerline{\ifx\picnaturalsize N\epsfxsize \picsize\fi \epsfbox{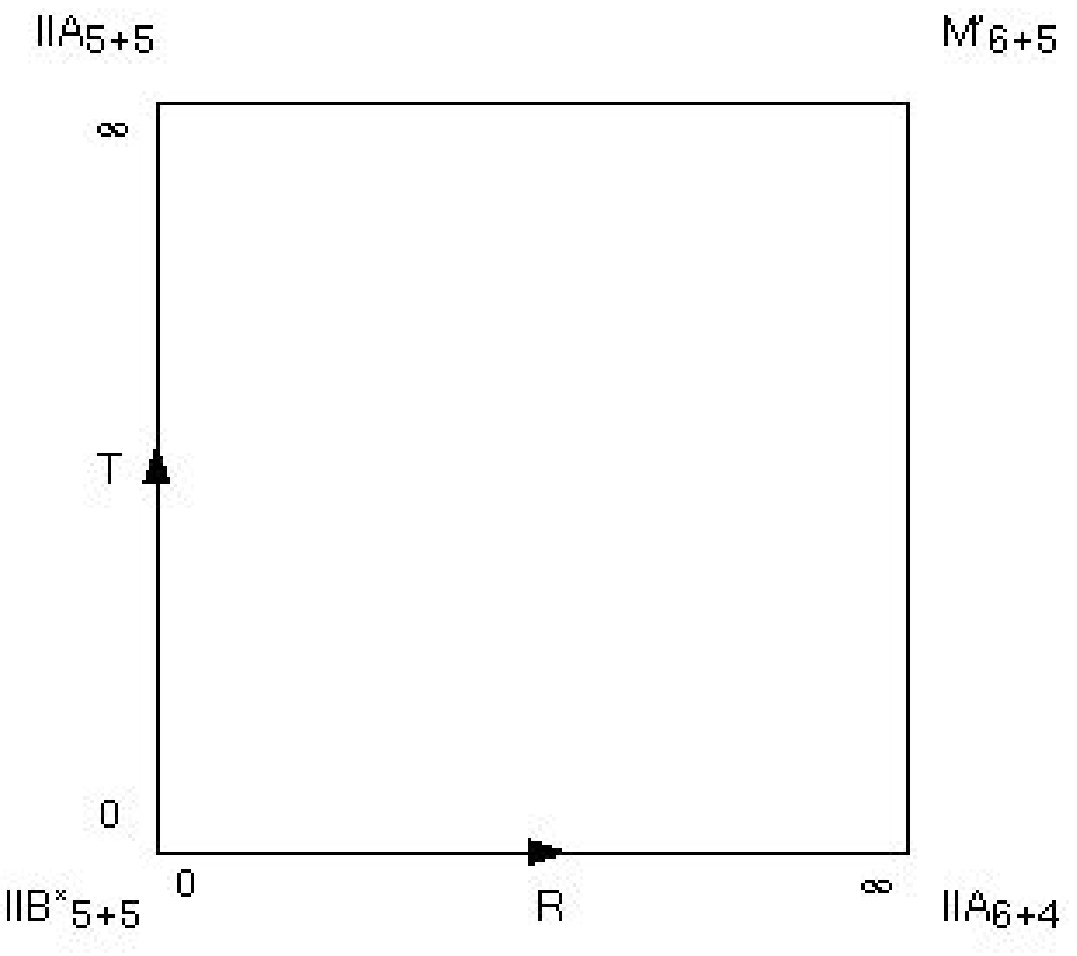}}}\fi
%%End InstantTeX Picture

{\bf Figure 10}  The moduli space for the compactification of   $M'$-theory from $6+5$  dimensions  on a  Lorentzian torus $T^{1,1}$  with  spacelike radius $R$ and  timelike radius  $T$.}}

\vskip .5 cm

{\vbox{

%%Begin InstantTeX Picture
\let\picnaturalsize=N
\def\picsize{3.0in}
\def\picfilename{SigFigs/fig8.eps}
%If you do not have the picture file add:
%\let\nopictures=Y
%to the beginning of the file.
\ifx\nopictures Y\else{\ifx\epsfloaded Y\else\input epsf \fi
\let\epsfloaded=Y
\centerline{\ifx\picnaturalsize N\epsfxsize \picsize\fi \epsfbox{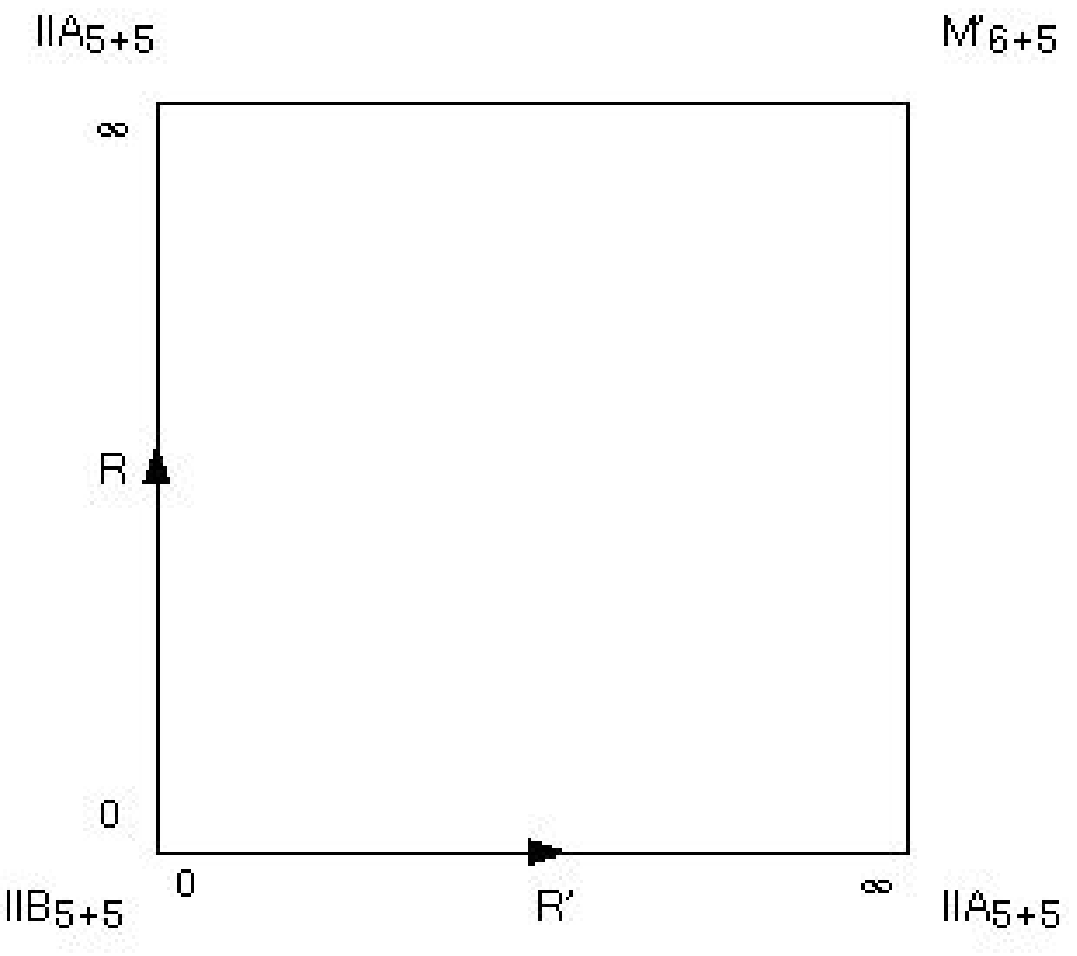}}}\fi
%%End InstantTeX Picture

 {\bf Figure 11}  The moduli space for the compactification of  $M'$-theory from $6+5$  dimensions    on a 
Euclidean torus $T^{2,0}$ with two spacelike circles of radii $R,R'$. The figure should be identified under reflection
 in the $R=R'$ diagonal.
}}

\vskip .5 cm

In the limits in which the torus $T^{2,0}$, $T^{1,1} $ or $T^{0,2}$ shrinks to zero size, we get the $IIB_{5+5}$,
$IIB^*_{5+5}$ or $IIB_{7+3}$
theories, respectively.

To understand the remaining 5+5 dimensional T-dualities in figure 4 from 11 dimensions, one needs 
to consider toroidal compactifications of the mirror theory in 5+6 dimensions, which is the strong
coupling limit of the type $IIA^*_{5+5}$ theory. These are depicted in figures 12,13,14.

\vskip 2 cm

{\vbox{

%%Begin InstantTeX Picture
\let\picnaturalsize=N
\def\picsize{3.0in}
\def\picfilename{SigFigs/fig9.eps}
%If you do not have the picture file add:
%\let\nopictures=Y
%to the beginning of the file.
\ifx\nopictures Y\else{\ifx\epsfloaded Y\else\input epsf \fi
\let\epsfloaded=Y
\centerline{\ifx\picnaturalsize N\epsfxsize \picsize\fi \epsfbox{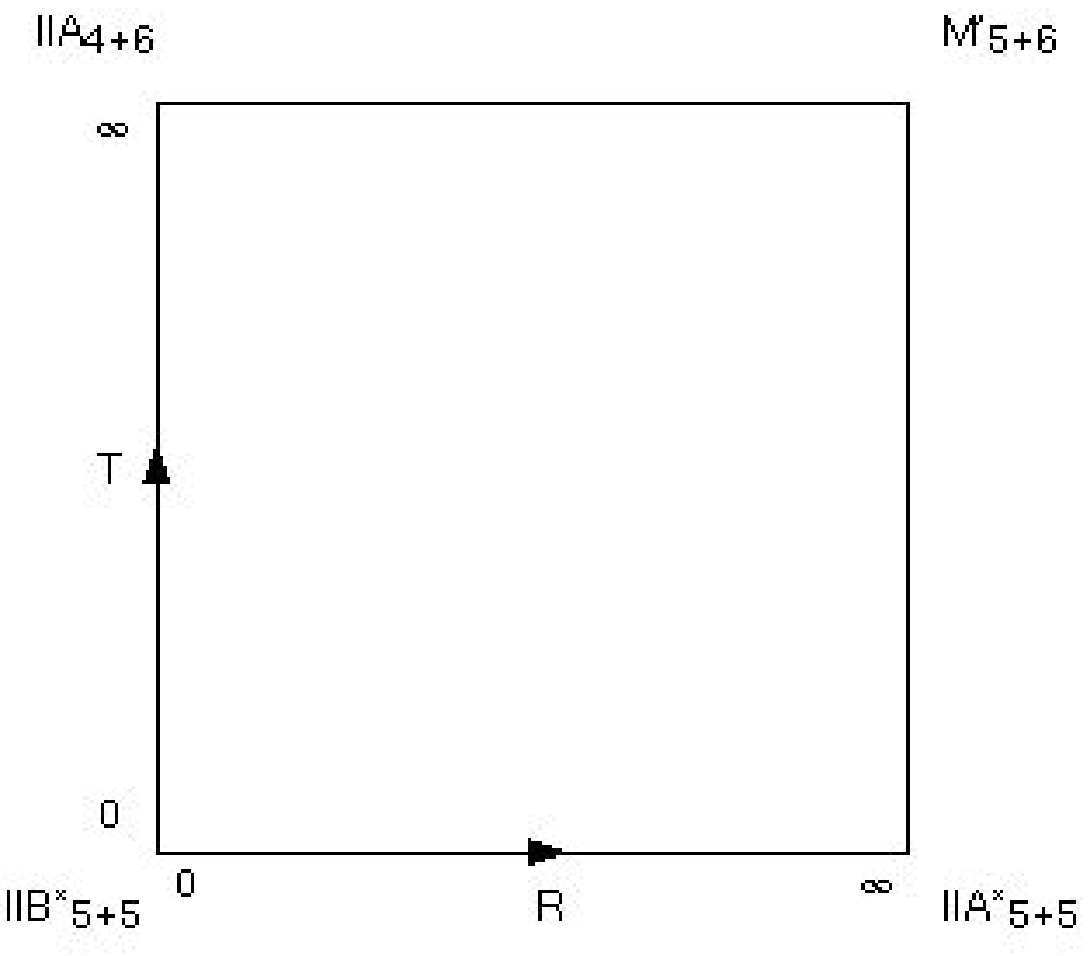}}}\fi
%%End InstantTeX Picture

{\bf Figure 12}  The moduli space for the compactification of   $M'$-theory from $5+6$  dimensions  on a  Lorentzian torus $T^{1,1}$  with  spacelike radius $R$ and  timelike radius  $T$.}}

%\vskip .5 cm

Figure 12 gives an $M'$-theory understanding of the  
 spacelike T-duality between the $IIA^*_{5+5}$ theory and the 
$IIB^*_{5+5}$ theory,
while figure 13 leads to the timelike T-duality between the $IIA^*+_{5+5}$ theory and the 
$IIB+_{5+5}$ theory.
Figure 14 leads to the signature-changing T-dualities between the $IIB_{3+7} $ and $IIA_{4+6}$ theories.

\vskip .5 cm

{\vbox{

%%Begin InstantTeX Picture
\let\picnaturalsize=N
\def\picsize{3.0in}
\def\picfilename{SigFigs/fig10.eps}
%If you do not have the picture file add:
%\let\nopictures=Y
%to the beginning of the file.
\ifx\nopictures Y\else{\ifx\epsfloaded Y\else\input epsf \fi
\let\epsfloaded=Y
\centerline{\ifx\picnaturalsize N\epsfxsize \picsize\fi \epsfbox{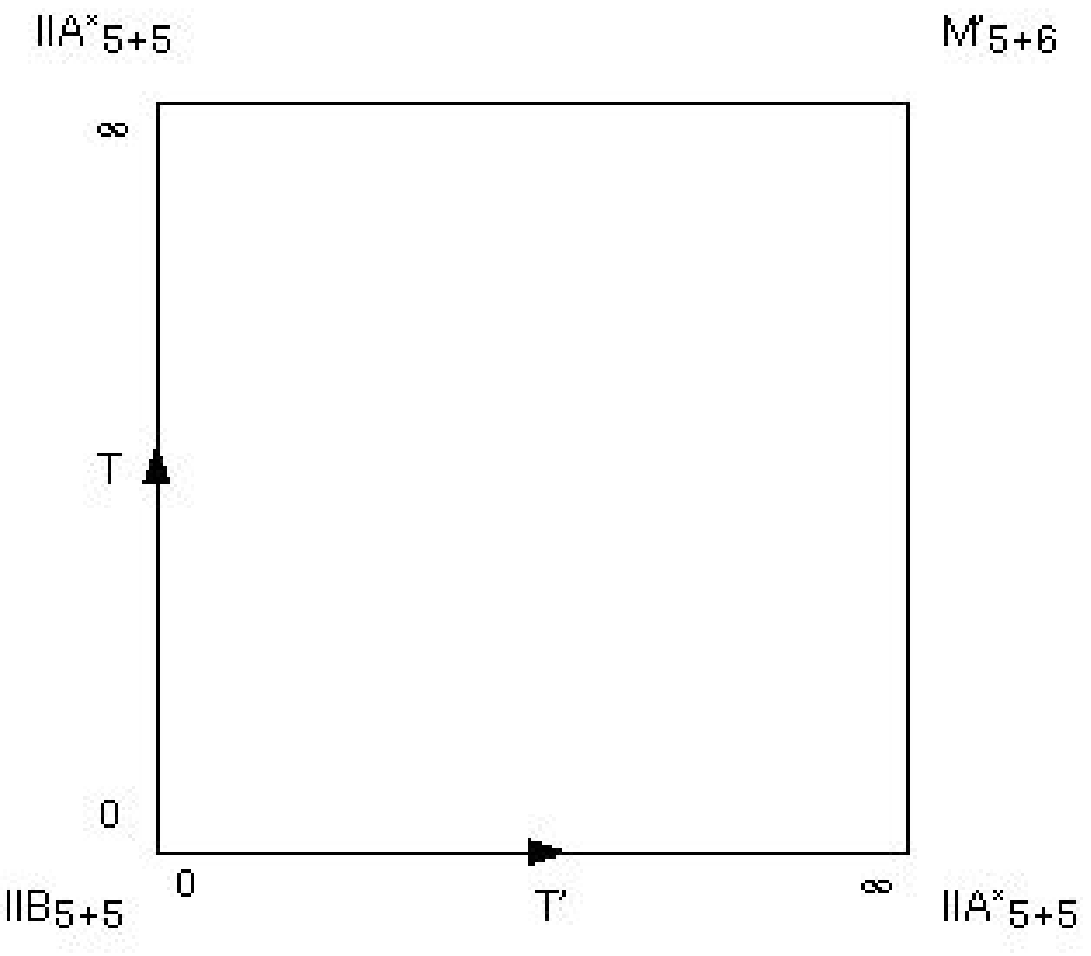}}}\fi
%%End InstantTeX Picture

{\bf Figure 13}  The moduli space for the compactification of $M'$-theory from $5+6$ dimensions on a
 torus $T^{0,2}$ with two timelike circles of radii $T,T'$. The figure should be identified under reflection in the $T=T'$ diagonal.
}}

\vskip .5 cm

{\vbox{

%%Begin InstantTeX Picture
\let\picnaturalsize=N
\def\picsize{3.0in}
\def\picfilename{SigFigs/fig11.eps}
%If you do not have the picture file add:
%\let\nopictures=Y
%to the beginning of the file.
\ifx\nopictures Y\else{\ifx\epsfloaded Y\else\input epsf \fi
\let\epsfloaded=Y
\centerline{\ifx\picnaturalsize N\epsfxsize \picsize\fi \epsfbox{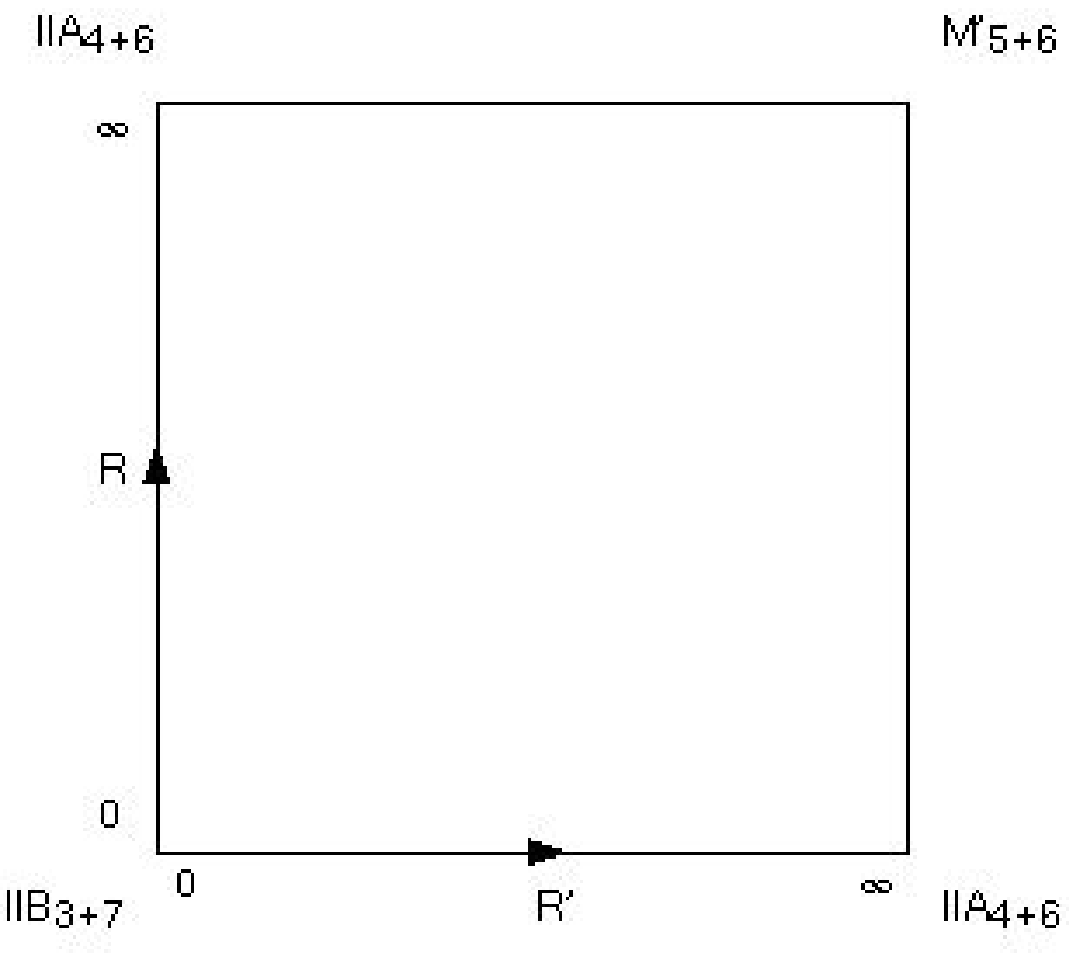}}}\fi
%%End InstantTeX Picture

{\bf Figure 14}  The moduli space for the compactification of  $M'$-theory from $5+6$  dimensions    on a Euclidean 
torus $T^{2,0}$ with two spacelike circles of radii $R,R'$. The figure should be identified under reflection in the $R=R'$
diagonal. }}

\vskip .5 cm

\chapter{Compactification on 3-Tori}

In this section we will consider the   M-theories of various signatures  compactified  
on 3-tori of various signatures, giving dualities linking the   M-theories.
M-theory in 10+1 dimensions compactified on $T^{2,0}$ gives the IIB string in the limit in which the
torus shrinks, while when compactified on a shrinking Lorentzian torus $T^{1,1}$ it gives the
$IIB^*$  string, as we have seen, so that a shrinking $T^{2,0}$ gives an extra spatial  coordinate (in addition to the 8+1 dimensions remaining after compactification), 
while a shrinking
$T^{1,1}$ gives an extra time  dimension (in addition to the 9+0 dimensions remaining after compactification).
The extra dimension  opens up because an infinite tower of states from the membrane wrapping modes contribute to the
massless spectrum in the limit, and these can be associated with the Kaluza-Klein modes from an extra dimension that is
decompactifying.
Then M-theory on $T^{3,0}$ gives 3 extra spatial dimensions in the shrinking limit from membranes wrapping around the three
2-cycles, so that these replace the three spatial dimensions lost in the compactification and a
T-dual  M-theory in 10+1 non-compact dimensions is regained in the limit. On $T^{2,1}$, there are 2
Lorentzian   $T^{1,1}$ cycles and one Euclidean
$T^{2,0}$ cycle and so, in the limit in which all three circles shrink, the 2 spatial and one time 
dimensions that are lost are replaced by two time and one space dimensions, so that the $M^*$ theory
in
$9+2$ dimensions is regained.

Similar arguments can be applied to the $M^*$ and $M'$ theories.
From the dualities in section 8, we see that $M^*_{9+2}$
theory compactified on a shrinking $T^{2,0}$ gives an extra time  dimension, on a shrinking $T^{1,1}$ gives an extra spatial 
dimension and 
on a shrinking $T^{0,2}$ gives an extra time 
dimension. Then $M^*_{9+2}$
theory compactified on a shrinking $T^{3,0}$ gives 6+2 dimensions plus 3 extra time dimensions to give a theory in 6 space
and 5 time dimensions, the $M'_{6+5}$ theory.
The $M^*_{9+2}$
theory compactified on a shrinking $T^{2,1}$ gives 7+1 dimensions plus an  extra two space and one time dimensions to give a
theory in 9 space and 2 time dimensions, which is the  $M^*_{9+2}$ theory again.
The $M^*_{9+2}$
theory compactified on a shrinking $T^{1,2}$ gives 8+0 dimensions plus an  extra two space and one time dimensions to give a
theory in 10 space and 1 time dimensions, which is the   original $M _{10+1}$ theory.

From the dualities in section 9, we see that $M'_{6+5}$
theory compactified on a shrinking $T^{2,0}$ gives an extra spatial  dimension, on a shrinking $T^{1,1}$ gives an extra time  
dimension and 
on a shrinking $T^{0,2}$ gives an extra spatial 
dimension. Then $M'_{6+5}$
theory compactified on a shrinking $T^{3,0}$ gives 3+5 dimensions plus 3 extra space dimensions to give a theory in 6 space
and 5 time dimensions, the $M'_{6+5}$ theory again.
The $M'_{6+5}$
theory compactified on a shrinking $T^{2,1}$ gives 4+4 dimensions plus an  extra two time and one  space dimensions to give a
theory in 5 space and 6 time dimensions, which is the  $M'_{5+6}$ theory.
The $M'_{6+5}$
theory compactified on a shrinking $T^{1,2}$ gives 5+3 dimensions plus an  extra two time and one  space dimensions to give a
theory in 6 space and 5 time dimensions, which is the     $M'_{6+5}$  theory.
Finally, $M'_{6+5}$
theory compactified on a shrinking $T^{0,3}$ gives 6+2 dimensions plus 3 extra space dimensions to give a theory in 9 space
and 2 time dimensions, the $M^*_{9+2}$ theory.

The torus reductions of the reversed signature cases are given by reversing all the signatures in the above.
Then we have a chain of dualities linking the \lq different' M-theories.
$M _{10+1}$ theory on a shrinking $T^{2,1}$ gives the $M^*_{9+2}$ theory, the $M^*_{9+2}$ theory on a shrinking 
$T^{3,0}$ gives the $M'_{6+5}$ theory, the $M'_{6+5}$ on a shrinking $T^{2,1}$ gives the $M'_{5+6}$
theory, the $M'_{5+6}$ theory on a shrinking 
$T^{3,0}$ gives the $M^*_{2+9}$ theory and the $M^*_{2+9}$ theory on a shrinking $T^{2,1}$ gives the
$M_{1+10}$ theory.
The usual 10+1 dimensional M-theory compactified on
 $T^{2,1}$ is equivalent to $M^*_{9+2}$ theory compactified on a dual  $T^{1,2}$,
$M^*_{9+2}$ theory compactified on
 $T^{3,0}$ is equivalent to $M'_{6+5}$ theory compactified on a dual  $T^{0,3}$,
$M'_{6+5}$ theory compactified on
 $T^{2,1}$ is equivalent to $M'_{5+6}$ theory compactified on a dual  $T^{1,2}$,
and so on, where in each case, the dual torus has a reversed signature as well as the inverse volume.
There are also self-dualities that preserve the signature, so that
for example the $M^*_{9+2}$ theory compactified on
 $T^{2,1}$ is equivalent to the same theory
  compactified on a dual  $T^{2,1}$.

\chapter{Supersymmetry and Fermions}

The dualities linking the various theories obtained in previous sections   guarantee that each of
the theories has 32 supersymmetries, but it is interesting to check how the details work out.
It will be useful to review  the types of spinors that can occur for each spacetime signature [\KT,\vann],
following [\KT]. In $D=s+t$ dimensions   with $s$ space dimensions and $t$ time
dimensions, the Clifford algebra is
$$\{ \ggg^m , \ggg ^n\} = -2 \eta^{mn}
\eqn\abc$$
where
$$ \eta ^{mn} = diag ( -1_t , 1_s)
\eqn\abc$$
and $m,n$ are tangent space indices.
The gamma matrices can be chosen to   satisfy
$$
\ggg_m ^\dagger =-(-1)^tA \ggg_mA^{-1}
\eqn\abc$$
where $A$ is the product of the timelike gamma matrices
$$ A=\ggg_1\ggg_2\dots \ggg_t
\eqn\abc$$
For even $D$,  $\pm \ggg_m^*$ constitute an equivalent representation of the Clifford algebra so
that there is a matrix $B$ such that
$$
\ggg_m ^* =\eta B\ggg_mB^{-1}, \qq \eta = \pm 1
\eqn\bcon$$
for either choice of $\eta$, while for odd $D$ there is a matrix $B$ satisfying \bcon\
only for
 $$\eta =  (-1)^{(s+1-t)/2}
\eqn\abc$$
The matrix $B$ is unitary and satisfies
$$B^*B=\ee (\eta,s,t), \qq  \ee=\pm 1
\eqn\abc$$
where $\ee$ depends on the choice of $\eta$ (if $D$ even) and on the signature $s,t$.
The spacetime signatures $(s,t)$ allowing the four possible combinations $(\ee,\eta)$ are [\KT]
$$
\eqalign{
 & \ee= +1, \eta =-1 : \qq s-t =0,1,2 ~ mod ~ 8 \cr
& \ee= +1, \eta =+1 : \qq s-t =0,6,7 ~ mod ~ 8 \cr
& \ee= -1, \eta =-1 : \qq s-t =4,5,6 ~ mod ~ 8 \cr
& \ee= -1, \eta =+1 : \qq s-t =2,3,4 ~ mod ~ 8 \cr
 }\eqn\abc$$

The charge conjugation matrix is 
$$
C=B^tA
\eqn\abc$$
and
$$
\ggg_m ^t =-(-1)^{t}\eta C \ggg_mC^{-1}
\eqn\abc$$
For even $D$, the matrix
$$\ggg_{D+1}= \ggg_1\ggg_2\dots \ggg_D
\eqn\abc$$
satisfies 
$$(\ggg_{D+1})^2=(-1)^{(s-t)
/2}
\eqn\abc$$
and  the chiral projection operators
$$
P_\pm=\2 \left( 1 \pm (-1)^{(s-t)
/4}\ggg_{D+1}\right)
\eqn\wey$$
  can be used to decompose a spinor into two Weyl spinors $\psi_\pm = P_\pm \psi$.

If $\ee=1$, then it is consistent to impose the reality constraint
$$
\psi^*=B\psi, \qq \ee=1
\eqn\maj$$ on spinors.
Spinors satisfying this are Majorana spinors if $\eta=-1$ and pseudo-Majorana spinors if $\eta=1$.
The two cases are distinct in general: for example, a Majorana fermion can be massive, but 
a pseudo-Majorana fermion cannot satisfy a massive Dirac equation
and so must be massless [\KT]. 
The Dirac conjugate of a spinor $\psi$ is
$$\bar \psi = \psi ^\dagger A
\eqn\abc$$
so a (pseudo-) Majorana spinor satisfies
$$\bar \psi = \psi ^t C
\eqn\abc$$
If $\ee=-1$, the condition \maj\ is inconsistent as $B^*B=-1$, but if there are two spinors $\psi _i$, $i=1,2$,
one can impose the condition
$$
 \psi^{*i}=(\psi _i)^*=\eee^{ij}B\psi_j
\eqn\sym$$
where 
$\eee ^{ij}$ is the alternating tensor.
Spinors satisfying this with $\eta=-1$ are symplectic Majorana spinors, while those satisfying this for $\eta =1$ are
symplectic pseudo-Majorana spinors.
Each of these four conditions is consistent with the chiral projection using \wey\ only if $s-t=0~ mod ~4$, in which case the
resulting spinor is (symplectic) (pseudo) Majorana-Weyl.

The condition \maj\ can be generalised to the condition 
$$
 \psi^{*i}=(\psi _i)^*=M^{ij}B\psi_j, \qq \ee =1
\eqn\mmaj$$
for $N$ spinors, $i=1,\dots , j$,  where $M^{ij}$ is a symmetric real matrix satisfying $M^2=1$, so that $M$ is invariant under
$O(p,q)$ for some $p,q$ where
$p+q=N$ is the number of spinors, and the condition \sym\ is invariant under the action of $O(p,q)$. 
We will refer to \mmaj\ as an $O(p,q)$-Majorana condition.
If $\ee=-1$ and there are $2N$ spinors, 
one can impose the condition
$$
 \psi^{*i}=(\psi _i)^*=\www^{ij}B\psi_j
\eqn\abc$$
for some antisymmetric real matrix $\www^{ij}$ satisfying $\www ^2=-1$, and this is invariant under
$Sp(2N)$.

%%%%

It is often possible to choose a Majorana representation of the gamma-matrices in which 
the gamma matrices are all real or all imaginary, so that  one   can   take  $B=1$ so that a (pseudo-) Majorana fermion is
real, $\psi ^*=\psi$. For example, in 9+1 dimensions, the  
gamma matrices can be chosen to be all imaginary,  $\ggg ^*_m = - \ggg_m$,
and there are two choices of $B$, depending on the sign of $\eta$. For
 $\eta =-1,\ee=1  $, one   can   take  $B=1$ so that a Majorana fermion is real, $\psi ^*=\psi$.
For  $\eta=1, \ee = 1$, one can instead choose $B= \ggg_{11}$ and a pseudo-Majorana spinor
satisfies $\psi ^*=\ggg_{11}\psi$.
Then Majorana-Weyl spinors satisfy $\psi _\pm ^* = \psi
_\pm$ while pseudo-Majorana-Weyl spinors satisfy $\psi _\pm ^* = \pm\psi _\pm$, so that a negative chirality
pseudo-Majorana-Weyl spinor is imaginary, while a positive chirality pseudo-Majorana-Weyl
spinor is real. Then $\psi _- \to i \psi _-$ takes a  Majorana or Majorana-Weyl spinor to a
pseudo-Majorana or pseudo-Majorana-Weyl spinor, and vice versa, so that Majorana and 
pseudo-Majorana conditions are equivalent in this sense, but the multiplication by $i$ changes the sign of the kinetic term of 
a $\psi_-$ fermion.
In particular, right-handed Majorana-Weyl and right-handed pseudo-Majorana-Weyl spinors
are both real and so equivalent; in 10 dimensions, this applies to signatures 9+1,5+5,1+9.
Similarly, 
right-handed symplectic Majorana-Weyl spinors and right-handed symplectic pseudo-Majorana-Weyl spinors are equivalent;
in 10 dimensions, this applies to signatures 7+3,3+7.
In signature (1,9),  all the gamma-matrices can be chosen to be real (they are $i$ times the imaginary gamma matrices of the Majorana reperesentation in signature 9+1)
so that $B=1$ for $\eta=1$ and pseudo-Majorana spinors are real.

Given two Majorana-Weyl spinors $\psi ^\pm$ of opposite chirality, they can be combined to form a Majorana spinor
$\Psi=(\psi^+,\psi^-)$ or a pseudo-Majorana spinor
$\Psi=(\psi^+,i\psi^-)$. The IIA and $IIA^*$ supergravity theories in 9+1 dimensions both have two gravitini $\psi_\mm ^+,
\psi_\mm^-$ which are Majorana-Weyl spinors of opposite chirality.
In the IIA supergravity, their kinetic terms have the same sign and can be written in terms of a Majorana gravitino
as
$$ \int  d^{10} x \sqrt
 g  \, i   \bar \psi   _\mu \Gamma^{\mu \nu \rho} \nabla _\nu \psi_\rho 
\eqn\ertada$$
or in terms of a pseudo-Majorana gravitino as
$$ \int  d^{10} x \sqrt
 g  \, i   \bar \psi   _\mu \Gamma^{\mu \nu \rho} \Gamma^{11}\nabla _\nu \psi_\rho 
\eqn\ertdsad$$
In the $IIA^*$ supergravity, the  kinetic terms of the two Majorana-Weyl gravitini
have opposite sign and can be written in terms of a Majorana
gravitino as \ertdsad\ or in terms of a pseudo-Majorana gravitino as
\ertada. 
In 10 dimensions, similar considerations apply to signatures  5+5,1+9
in which there are also   Majorana-Weyl spinors. 

Two Majorana spinors $\psi, \chi$ can be combined into an $SO(2)$-Majorana spinor $\psi^i=(\psi, \chi)$, $i=1,2$,
satisfying
\mmaj\ with $M^{ij}=\delta ^{ij}$, or into an 
$SO(1,1)$ Majorana spinor $\psi^i=(\psi, i\chi)$  satisfying \mmaj\ with $M^{ij}=\eta ^{ij}$, and similarly for
pseudo-Majorana spinors or
 Majorana-Weyl
spinors of the same chirality. The     type IIB and type $IIB^*$ supergravities both  have two Majorana-Weyl gravitini   of 
the same chirality 
and the signs of their kinetic terms are the same for the  type IIB  supergravity and opposite for the
type $IIB^*$ supergravity.
In the type IIB theory, the gravitini can be combined into an $SO(2)$-Majorana-Weyl gravitino $\psi ^i_\mm$
with kinetic term  
$$ \int  d^{10} x \sqrt
 g  \, i  \delta_{ij}\bar \psi ^i _\mu\Gamma^{\mu \nu \rho} \nabla _\nu \psi_\rho^j
\eqn\asdfr$$
or into an   $SO(1,1)$-Majorana-Weyl gravitino $\psi ^i_\mm$
with kinetic term  
$$ \int  d^{10} x \sqrt
 g  \, i  \eta_{ij}\bar \psi ^i _\mu\Gamma^{\mu \nu \rho} \nabla _\nu \psi_\rho^j
\eqn\asddfgdffr$$
The two Majorana-Weyl gravitini of the
type $IIB^*_{9+1}$ theory can   
be combined into either an 
$SO(2)$-Majorana-Weyl gravitino $\psi ^i_\mm$
with kinetic term  \asddfgdffr\ or into an
$SO(1,1)$-Majorana-Weyl gravitino $\psi ^i_\mm$
with kinetic term  \asdfr.

%%%%%%%
For every theory in signature $(s,t)$ we have found a mirror theory with signature
$(t,s)$. The bosonic part of the supergravity actions of the mirror pairs are equivalent (up to an
overall sign), but there appear to be differences in some cases in the fermionic sectors.
For example, there are Majorana fermions in signature 10+1 but not in signature 1+10, where, from above, only
pseudo-Majorana fermions are allowed.
In 11 dimensions, the fermions are Majorana in signatures (10,1),(6,5) and (2,9) and are pseudo-Majorana
in the mirror signatures (1,10),(5,6) and (9,2), 
and in each case the spinors have 32 real components.
However,
a Majorana spinor in 10+1 and a pseudo-Majorana spinor in 1+10 dimensions are both real in a Majorana representation, so    that the two mirror theories are equivalent, and similarly
for other mirror pairs.

Indeed, consider a 
theory in signature $(s,t) $
  with  the spinors satisfying a
Majorana-type condtion $\psi ^* = NB\psi $ (which can be pseudo or symplectic)
 for a   $B$-matrix with corresponding $\epsilon,\eta$,
and for some matrix $N$ which is the identity for   (pseudo)-Majorana spinors and is antisymmetric $N=\Omega$
for symplectic  (pseudo)-Majorana spinors.
Taking $g_{\mu \nu} \to -g_{\mu\nu}$, $\Gamma_ \mu  \to i \Gamma _\mu$,
gives a theory in signature $(t,s)$ in which the spinors 
still satisfy $\psi ^* = NB\psi $,   
  but $\eta $ has changed sign, so that
a  (symplectic) Majorana spinor   changes   to (symplectic)  pseudo-Majorana, or vice versa. This mirror theory will also be supersymmetric, and so must be the same as the mirror obtained via the chain of dualities, as supersymmetry specifies the theory uniquely.
Thus the mirror pairs of theories are all equivalent.

Dimensionally reducing to 10 dimensions on either a spacelike or a timelike circle, we obtain IIA theories in which the
spinor is a 32-component Majorana or pseudo-Majorana spinor, depending on the signature as follows:
$$\eqalign{
Majorana: \qq & IIA_{9+1},IIA_{10+0},IIA_{6+4},IIA_{5+5},IIA_{2+8},IIA_{1+9}^* \cr
Pseudo-Majorana: \qq & IIA^*_{9+1},IIA_{8+2},IIA_{5+5}^*,IIA_{4+6},IIA_{1+9},IIA_{0+10} \cr}
\eqn\abc$$
In signatures $9+1,5+5,1+9$, the spinors can be decomposed into   Majorana-Weyl spinors.
Note that    two kinds of spinors are possible for each signature (e.g. 9+1 dimensions allows Majorana  or
pseudo-Majorana  spinors), but the relation with 11 dimensions picks out one type  in each case.  However, in signatures (9,1), (1,9) and (5,5), the pseudo-Majorana spinors can be repackaged as Majorana spinors, as seen above.

The 11-dimensional superalgebra for all of the signatures is of the form
$$\{ Q,Q\} = ( \ggg_M C^{-1}) P^M
\eqn\abcty$$
where $C$ is the $D=11$ charge conjugation matrix and the supercharges $Q$ are either Majorana or pseudo-Majorana, depending
on the signature. A spatial dimensional reduction will give a superalgebra of the same form in $D=10$, while for a  reduction
on a timelike direction $x^{11}$, the 10-dimensional charge conjugation matrix $C_{10}$ is related to that in 11-dimensions by
$C=C_{10}\ggg_{11}$, so that the 10-dimensional algebra is of the twisted form
$$\{ Q,Q\} = ( \ggg_M \ggg_{11} C_{10}^{-1}) P^M
\eqn\abchjkl$$
For signatures (9,1),(5,5) and (1,9) this can be rewritten in terms of two
(pseudo)-Majorana-Weyl supercharges $Q_i$, $i=1,2$, of opposite chirality, in which case it becomes
$$\{ Q_i,Q_j \} = \eta_{ij}( \ggg_M C^{-1}) P^M
\eqn\abchjgf$$
where $\eta_{ij}=diag (+1,-1)$, so that \salg\ is recovered. This is a general feature: each timelike reduction introduces  a
twist into the superalgebra.
In 9+1,5+5 or 1+9 dimensions, the 
Majorana-Weyl supercharges $Q_i$ can be combined into a Majorana or a pseudo-Majorana supercharge, and
 the twisted superalgebra
\abchjgf\ can be written as \abchjkl\ in terms of a Majorana supercharge or as 
\abcty\ in terms of a pseudo-Majorana supercharge.

For the IIB theories in signatures 9+1,5+5,1+9, there are two Majorana-Weyl supercharges $Q_i$ of the same chirality which 
satisfy an untwisted superalgebra \untw\
  for the
$IIB_{9+1},IIB_{5+5},IIB_{1+9}$ theories, or a twisted superalgebra
 for the
$IIB_{9+1}^*,IIB_{5+5}^*,IIB_{1+9}^* $ theories. (These can be rewritten in terms of either an $SO(2)$-Majorana-Weyl
supercharge or an $SO(1,1)$-Majorana-Weyl supercharge.)
 For the $IIB_{7+3}$ theory and the the $IIB_{3+7}$ the  supercharges $Q_i$, $i=1,2$, form a symplectic Majorana-Weyl spinor,
satisfying
\sym, and the superalgebra takes the symplectic twisted form
$$\{ Q_i,Q_j \} = \ee_{ij}( \ggg_M C^{-1}) P^M
\eqn\abc$$

\chapter{The Set of Theories}

We have found versions of M-theory   with signatures $(10,1),(9,2),(6,5)$ and the mirrors
 $(1,10),(2,9),(5,6)$, and  these   are all linked by dualities -- we saw in  section 10 that we could get from one to another by
compactifying on a 3-torus of suitable signature and taking the limit in which it shrinks to zero
size. All can be linked by dualities in this way, so that they should not be regarded as different
theories, but as  different limits of the same underlying theory.
All have     field theory limits which are 11-dimensional supergravity theories with various
signatures. 
 Compactifying on
a spacelike circle or a timelike circle gives IIA string theories with signatures
$(10,0),(9,1),(8,2),(6,4),(5,5)$ together with their mirrors, and the starred versions of the theories with
signature $(9,1)$, $(1,9)$ and $(5,5)$ in which the signs of the RR fields are reversed. 
Compactifying the various M-theories  on shrinking 2-tori $T^{2,0}$, $T^{1,1} $ or $T^{0,2}$ give IIB theories in
signatures $(9,1), (7,3),(5,5)$ together with their mirrors, and the starred versions of the theories with
signature $(9,1)$, $(1,9)$ and $(5,5)$.
The strong coupling limit of the $IIB^*$ theories gives the $IIB'$ theories, while the remaining $IIB$ theories are self-dual,
 so that their strong coupling limit is isomorphic to the weak
coupling limit, as in the usual IIB theory [\HT].
The various 10-dimensional string theories are all linked by T-dualities or by S-dualities (taking
weak coupling to strong coupling), as shown in figure 4.

The various 11-dimensional supergravities have a   field content of a metric, a 3-form gauge field
and a gravitino which is Majorana for signatures $10+1,6+5,2+9$ and pseudo-Majorana for the mirror
signatures $1+10,5+6,9+2$. The bosonic Lagrangian is of the form
$R-(dC_3)^2+\dots$ for  signatures $10+1,6+5,2+9$ while the 3-form gauge field kinetic term has the
wrong sign for the signatures $1+10,5+6,9+2$, with lagrangian 
$R+(dC_3)^2+\dots$.

The bosonic fields of the type IIA supergravities are the NS-NS fields $g_{\mm \nn},B_2,\Phi$ and the RR
gauge fields $C_1,C_3$ with a bosonic action of the form
$$\eqalign{
S_{IIA}= 
\int
 d^{10} x &
\sqrt{\vert g\vert }\left[
 e^{-2 \Phi} \left( R+ 4(\partial   \Phi )^2  
\right)\right.
\cr &
\left. -\left( \pm e^{-2 \Phi} H^2 
\pm G_2^2 \pm G_4^2 \right)\right] +   \dots
\cr}
\eqn\twoapm$$
with 
the signs of the kinetic terms of $B_2,C_1,C_3$ for the various spacetime signatures  
given in table 1 (where a $+$ sign is the usual sign, and   $-$ is the sign for a ghost-like field).

\vskip 0.5cm
\begintable
 Theory | $C_1$ Sign | $B_2$ Sign | $C_3$ Sign | Spinor \elt
 $IIA_{9+1}$ | + | + | + | M=MW+MW' \elt
 $IIA_{9+1}^*$ | - | + | - | PM=MW+MW' \elt
 $IIA_{10+0}$ | - | - | + | M \elt
 $IIA_{8+2}$ | + | - | - | PM \elt
 $IIA_{6+4}$ | - | - | + | M \elt
 $IIA_{5+5}$ | + | + | + | M=MW+MW' \elt
 $IIA_{5+5}^*$ | - | + | - | PM=MW+MW' \elt
 $IIA_{4+6}$ | + | - | - | PM \elt
 $IIA_{2+8}$ | - | - | + | M \elt
 $IIA_{1+9}^*$ | + | + | + | M=MW+MW' \elt
 $IIA_{1+9}$ | - | + | - | PM=MW+MW' \elt
 $IIA_{0+10}$ | + | - | - | PM 
\endtable
 
{{\bf Table 1} The signs of the kinetic terms in \twoapm\ for the gauge fields $C_1,B_2,C_3$ for the $IIA_{s+t}$ theory 
in $s$ space and $t=10-s$ time dimensions. A \lq +' sign is the conventional sign. The spinors are Majorana (M) or
pseudo-Majorana (PM), and in certain cases split into positive chirality Majorana-Weyl   spinors (MW)  and negative chirality
ones (MW').}

\vskip .5cm

For the type IIB supergravities, the bosonic fields are the NS-NS fields $g_{\mm \nn},B_2,\Phi$ and
the RR fields $C_0,C_2,C_4$, where the $C_4$ field strength $G_5$ is self-dual, and the bosonic
action is of the form
$$\eqalign{
S_{IIB}= 
\int
 d^{10} x &
\sqrt{\vert g \vert }\left[
 e^{-2 \Phi} \left( R+ 4(\partial   \Phi )^2  
\right)\right.
\cr &
\left. -\left( \pm e^{-2 \Phi} H^2 \pm G_1^2
\pm G_3^2 \pm G_5^2 \right)\right] +   \dots
\cr}
\eqn\twobpm$$
with the
signs of the
kinetic terms of $B_2,C_0,C_2,C_4$ for the various spacetime signatures   given in table 2.

\vskip 0.5cm
\begintable
 Theory | $C_0$ Sign | $B_2$ Sign | $C_2$ Sign | $C_4$ Sign | Spinor \elt
 $IIB_{9+1}$ | + | + | + | + |  MW +MW \elt
 $IIB_{9+1}^*$ | - | + | - | - |  MW +MW \elt
 $IIB_{9+1}'$ | - | - | + | - | MW +MW \elt
 $IIB_{7+3}$ | + | - | - | + | SMW \elt
 $IIB_{5+5}$ | + | + | + | + | MW +MW \elt
 $IIB_{5+5}^*$ | - | + | - | - | MW +MW \elt
 $IIB_{5+5}'$ | - | - | + | - |MW +MW \elt
 $IIB_{3+7}$ | + | - | - | + | SMW \elt
 $IIB_{1+9}$ | + | + | + | + | MW +MW \elt
 $IIB_{1+9}^*$ | - | + | - | - | MW +MW \elt
 $IIB_{1+9}'$ | - | - | + | - | MW +MW 
\endtable
{{\bf Table 2} The signs of the kinetic terms in \twobpm\ for 
the gauge fields $C_0,B_2,C_2,C_4$ for the $IIB_{s+t}$ theory in $s$ space and $t=10-s$ 
time dimensions. A \lq +' sign is the
conventional sign. The spinors are positive chirality Majorana-Weyl (MW) or symplectic
Majorana-Weyl (SMW).}

\vskip .5cm

\noindent For the IIB theories, the scalar coset space is $SL(2,\R)/SO(2)$ when the $C_0$ sign is $+$ and 
is $SL(2,\R)/SO(1,1)$ when the $C_0$ sign is $-$.

Note that the theories are named so that the signature-reversing mirror 
transformation takes 
  the $IIA_{9+1}$ theory to the
 $IIA_{1+9}$ theory, 
  the $IIA_{9+1}^*$ theoryto the
 $IIA_{1+9}^*$ theory, 
  the $IIB_{9+1} $ theory to the
 $IIB_{1+9}$ theory, 
  the $IIB_{9+1}^* $  theory to the
 $IIB_{1+9}^*$ theory, and
 the $IIB_{9+1}' $  theory to the
 $IIB_{1+9}'$ theory.
The mirror transformation also takes the   $IIA_{5+5} $ theory to the
 $IIA_{5+5}^*$ theory, but leaves each of the three IIB theories in 5+5 dimensions invariant.

There are Majorana-Weyl spinors in signatures 9+1,1+9 and 5+5  and it
is only in these signatures in which $N=1$ theories  with 16 supersymmetries are possible, and it is in 
 these signatures that
supersymmetric Yang-Mills theories,  type I  strings or heterotic strings can occur.
Orientifolding the type $IIB$,$IIB^*$ and $IIB'$ theories in 9+1,5+5 and 1+9 dimensions gives type
$I$,$I^*$ and $I'$ theories in these signatures. The type I theories have both open strings and
D-strings with Lorentzian world-sheets, the type $I^*$ theories have open strings with Lorentzian 
world-sheets and D-strings with Euclidean world-sheets, while the
type $I'$ strings have Euclidean open strings and Lorentzian D-strings.
At strong coupling, the D-strings become, as in [\huten-\witpol],  the fundamental strings of the
corresponding dual heterotic  theories, which will be denoted $HO,HO^*,HO'$ respectively.
There are also  $HE$-type heterotic strings in these signatures that can be obtained by
compactifying $M,M^*,M'$ theories  on   a spacelike or a timelike $S^1/\Z_2$, when possible.
For example reducing M-theory on a spacelike $S^1/\Z_2$ gives the $HE_{9+1}$ theory, as in
[\horwit], reducing
$M^*_{9+2}$ theory on a timelike $S^1/\Z_2$ gives a $HE^*_{9+1}$ theory with fundamental strings
with Euclidean world-sheets,  reducing
$M'_{6+5}$ theory on a spacelike $S^1/\Z_2$ gives the $HE_{5+5}$ theory with Lorentzian
world-sheets, and reducing the 
$M'_{5+6}$ theory on a timelike $S^1/\Z_2$ gives the $HE^*_{5+5}$ theory with Euclidean
world-sheets.
The various heterotic strings are then related to each other by T-dualities, and to type II
theories by generalisations of the dualities in [\HT]. For example, the type $IIA_{5+5}$ theory
compactified on K3 is dual to the $HO_{5+5}$ or $HE_{5+5}$ heterotic theory compactified on $T^4$,
while the $IIA^*_{9+1}$ theory on K3 is dual to the $HO^*_{5+5}$ or $HE^*_{5+5}$ heterotic theory
on a timelike $T^4$.
Such dualities can be used to relate heterotic timelike T-duality to type II timelike T-duality.
 
Compactifications of type IIB theory on a space $K$ for which the complex scalar field $\tt$ is not
a function on
$K$ but is the section of a bundle can be described as a compactification of F-theory on a $T^2$
bundle $B$ over $K$  [\fvaf]. If $K$ is an $n$-dimensional Euclidean space, then $B$ is an
$n+2$-dimensional Euclidean space, and the F-theory background is $11+1$ dimensional. On the other hand, compactifications of type $IIB^*_{9+1}$ theory on
a space $K$ for which the   scalars taking values in $SL(2,\R)/SO(1,1)$ 
is  the section of a bundle
over
$K$ can be described as a compactification of an  $F^*$-theory on a $T^{1,1}$ bundle $B$ over $K$. If
$K$ is an
$n$-dimensional Euclidean space, then
 $B$ is an $n+2$-dimensional  space with Lorentzian signature, and the $F^*$-theory background is
$10+2$ dimensional, with two times [\hsta]. 
Similarly,  compactifications of the various $IIB$ theories in signatures
9+1,7+3,5+5 and their mirrors  in which the scalars   
have non-trivial transition functions
can be described as compactifications of F-type theories in signatures
11+1,10+2,9+3,7+5,6+6 and their mirrors.

\ack
{I would like to thank   Gary Gibbons,   Ramzi Khuri and Greg Moore for helpful discussions.}

\refout
\bye